\documentclass[pr,10pt,aps,twocolumn,showpacs,showkeys,tightenlines]{revtex4}
\usepackage{graphicx}

\begin{document}

\title{Regge phenomenology of pion photoproduction off the nucleon at forward angles}


\author{Byung Geel Yu}%

\email{bgyu@kau.ac.kr}%
\affiliation{Research Institute of Basic Sciences, Korea Aerospace
University, Koyang, 412-791, Korea}

\author{Tae Keun Choi}%

\email{tkchoi@yonsei.ac.kr}%
\affiliation{Department of Physics, Yonsei University, Wonju,
220-710, Korea}

\author{W. Kim}%

\email{wooyoung@knu.ac.kr}%
\affiliation{Department of Physics, Kyungpook National University,
Daegu, 702-701, Korea}

\date{\today}


\begin{abstract}

We present a Regge model for pion photoproduction which is
basically free of parameters within the framework of the
$s$-channel helicity amplitude. For completeness we take into
account axial mesons $a_1(1260)$, $b_1(1235)$ and tensor meson
$a_2(1320)$ in addition to the primary $\pi+\rho$ exchanges for
charged pion photoproduction, while the axial meson $h_1(1170)$
exchange is added to the model of $\omega+\rho^0+b_1$ exchanges
for the neutral case. The present model deals for the first time
with the $a_2$ and $h_1$ Regge poles in the $s$-channel helicity
amplitude. For model independence, we use coupling constants of
all exchanged mesons determined from empirical decay widths or
from the SU(3) relations together with consistency check with
existing estimates that are widely accepted in other reaction
processes. Based on these coupling constants the simultaneous
description of four photoproduction channels is given. Within the
Regge regime, $s\gg 4M^2$ and $-t < 2$ GeV$^2$, cross sections and
spin polarization asymmetries at various photon energies are
analyzed and results are obtained in better agreement with
experimental data without referring to any fitting procedure. The
model confirms dominance of the nucleon Born term in the sharp
rise of the charged pion cross section at very forward angles,
while dominance of the $\omega$ exchange with the nonsense wrong
signature zero leads to the deep dip in the neutral pion cross
section. In contrast to existing models, however, our model for
the charged pion case shows quite a different production mechanism
due to the crucial role of the tensor meson $a_2$ exchange in the
cross section and spin polarization asymmetries. Also the axial
meson $b_1$ exchange is found to give a sizable contribution to
the photon polarization asymmetry. In the neutral case, the role
of the $b_1$ is not significant, but the isoscalar $h_1$ exchange
gives an important contribution to the dip-generating mechanism in
the photon polarization, showing the isoscalar nature of the
process with the $\omega$. These findings demonstrate validity of
the present model with the prompt use of the tensor meson $a_2$
and axial meson $h_1$ for a wider application.

\end{abstract}

\pacs{13.40.-f, 13.60.Rj, 13.75.Jz, 13.88.+e}
\keywords{Pion
photoproduction, Regge pole, $s$-channel helicity amplitude,
Tensor meson, Axial meson}
\maketitle

\section{Introduction}

For decades progress has been made in both theory and experiment
to establish the model for the photoproduction of pseudoscalar
mesons from threshold to the resonance region. The effective
Lagrangian involving the chiral loop diagram has been applied to
realize the aspect of the soft pion dynamics \cite{chpt}, which
gradually extends to the intermediate energy region where the
low-lying resonance arises from a quasi-bound state of the
meson-baryon coupled channel \cite{ramo,geng}. These models have
been developed and tested along with the advent of the
experimental facilities, such as the SAPHIR/ELSA, the CLAS/JLab
and the LEPS/SPring-8 \cite{mami,unit,saph,clas,leps}.
But now, however, confronted with on-going plans in these
facilities to upgrade the energy scale higher than several tens of
GeVs \cite{exp}, the need for theoretical tools to describe the
dynamics of such production mechanisms at high energies that are
beyond the scope of the effective Lagrangian approach has grown.

It is known that the Regge formalism can serve to this end with
many unknown resonances in the $s$-channel replaced by the
representative $t$-channel Regge pole by duality \cite{fesr}. The
exchange of the spin quantum number carried by the meson
trajectory could provide a most economical way for the description
of high-energy phenomenologies
\cite{stro,worden,kell,sibi,sibi1,levy,guid,ghent,mosel}. In Ref.
\cite{kell}, Kellett considered the $\pi+\rho+a_2$ Regge pole
exchanges for the charged pion, and $\omega+\rho+b_1$ exchanges
for the neutral case together with their respective cuts in the
analysis of high-energy pion photoproduction. He obtained a good
fit of all available data on charged and neutral pion cases
simultaneously. More recently, Sibirtsev $et$ $al$. repeated the
same procedure based on an extended version of Ref. \cite{kell}.
With a global analysis of all world data available, they
investigated the applicability of the Regge approach to pion
photoproduction in the region $2\leq\sqrt{s}\leq 3$ GeV, the
so-called the fourth resonance region, by extrapolating the cross
section at high energy down to the resonance region for comparison
\cite{sibi,sibi1}.

Despite the apparent simplicity, however, these models
\cite{kell,sibi,sibi1} are based on the $t$-channel helicity
amplitude (TCHA) which totally requires a fit. On the other hand,
utilizing the photon helicity amplitude specific to the present
process \cite{walker}, Levy, Majerotto and Read (LMR) constructed
a model for the $K+K^*$ Regge pole exchanges in the photo and
electroproduction of kaon (and $\pi+\rho$ exchanges for the case
of pion) \cite{levy}. Since the $s$-channel helicity amplitude
(SCHA) of the case is given in terms of the conventional
Chew-Goldberger-Low-Nambu (CGLN) amplitude \cite{cgln} the
photoproduction current can be obtained by using the Born
approximation for the $t$-channel exchange of these mesons with
the couplings of the meson trajectories to photon or to baryons
implemented by the relevant interaction Lagrangians. It is,
therefore, advantageous to work with the Regge poles in the SCHA
in that one exploits the estimate from the decay width or from the
symmetry consideration for the coupling constants of the exchanged
meson in the use of the effective Lagrangians.
Furthermore, of the photoproduction current, gauge invariance is
easily prescribed for the exchange of the pion trajectory by
introducing the nucleon Born term, as Guidal, Laget, and
Vanderhaeghen(GLV) showed \cite{guid}. It turned out that the
nucleon Born term could account for the sharp rise of the charged
pion cross section at the very forward angle, thus removing the
theoretical uncertainty in this region due to the application of
the absorptive cut \cite{levy} or the parity-doublet conspiring
pion \cite{old,old1}.

At the present stage, however, the achievements of these models
are limited. The numerical consequences in the cross section and
spin polarizations show the deficiencies to explain the
experimental data at high energy and larger momentum transfer,
although the coupling constants of the $\rho$ exchange are taken
to be rather strong. In those versions extended to kaon
photoproduction \cite{levy,guid} this tendency becomes even
stronger to give the $K^*$ coupling constants too large values to
be realistic. This in turn may disprove that the models of
$\pi+\rho$, and of $K+K^*$ exchanges are too simple to be
realistic.
Meanwhile, the role of the tensor meson exchange, though not yet
considered in these models, is found to be of significance as the
natural parity exchange in the fit of the Regge poles using the
TCHA \cite{kell,sibi,sibi1}.

Motivated by such shortcomings in current model calculations,
i.e., the absence of the higher spin exchange and, as a
consequence, the poor parametrization of the meson coupling
constants, we here investigate the contribution of the higher spin
exchange based on the primary $\pi+\rho$ exchanges to search for
the possibility of the Regge approach to pion photoproduction
without fit parameters. In the present analysis of the process up
to the regime, $-t< 2$ GeV$^2$ and $s\gg 4M^2$, we include the
exchange of the $a_2(1320)$ meson in the charged pion case with a
particular attention to its role dissociated from the leading
$\rho$ trajectory. We are also interested in estimating the
contribution of the axial meson $h_1(1170)$ to the neutral case,
since these are the mesons to be investigated with their roles for
the first time  in the Regge model utilizing the SCHA. We expect
that the result of the present work could provide a reliable base
for the study of the resonance as found in recent approaches to
photo- and electroproduction processes \cite{ghent,mosel}.

This article is organized as follows. In Sec. II we begin with a
brief introduction of the SCHA for the present model calculation.
An extension of the Regge model follows to include the tensor
meson $a_2(1320)$ for the natural parity exchange. For the
unnatural parity exchange, we take into account the axial meson
$a_1(1260)$ and $b_1(1235)$ exchanges in the charged pion and the
$h_1(1170)$ exchange in the neutral pion case. Section III is
devoted to the determination of the coupling constant of the
exchanged meson prior to application. The radiative decay constant
of the exchanged meson is estimated either by using the measured
decay width or by the axiomatic meson-dominance hypothesis. The
strong coupling constant of the exchanged meson is determined from
the SU(3) relation. The numerical results with discussion are
presented in the Sec. IV. Three appendices follow with each part
containing materials for a more specific discussion.

\section{Regge context for pion Photoproduction}

For the Regge approach to pion photoproduction $\gamma(k)+N(p)\to
\pi(q)+N(p')$ in the SCHA, it is convenient to start with the four
positive photon helicity amplitudes defined by Walker
\cite{walker}
\begin{eqnarray}\label{scha-00}
&&H_1=-\frac{1}{\sqrt{2}}\,\sin\theta\,\cos\frac{\theta}{2}\,(F_3+F_4)\,,
\nonumber\\
&&H_2=-\frac{2}{\sqrt{2}}\,\cos\frac{\theta}{2}\,(F_1+F_2)+H_3\,,
\nonumber\\
&&H_3=\frac{1}{\sqrt{2}}\,\sin\theta\,\sin\frac{\theta}{2}\,(F_3-F_4)\,,
\nonumber\\
&&H_4=\frac{2}{\sqrt{2}}\,\sin\frac{\theta}{2}\,(F_1-F_2)-H_1\,,
\end{eqnarray}
where $\theta$ is the production angle between the photon and pion
three-momenta. The $F_i$ is the CGLN amplitude defined in the
nonrelativistic reduction of the photoproduction amplitude in the
center-of-mass frame \cite{cgln},
\begin{widetext}
\begin{eqnarray}
\frac{\sqrt{MM'}}{4\pi W}{\cal M }=F_1\,\sigma\cdot\hat\epsilon
+F_2\,i\,\sigma\cdot\hat{q}\,\sigma\cdot(\hat{k}\times{\hat\epsilon})
+F_3\,\sigma\cdot\hat{k}\,\hat{q}\cdot\hat{\epsilon}
+F_4\,\sigma\cdot\hat{q}\,\hat{q}\cdot{\hat\epsilon}\,,
\end{eqnarray}
\end{widetext}
with the photon polarization vector $\hat\epsilon$ and the
three-momenta $\hat{k}=\vec{k}/|\vec{k}|$ and
$\hat{q}=\vec{q}/|\vec{q}|$ for photon and pion. The $W$ is the
invariant energy of the system and the $M(M')$ is the mass of the
initial- (final-) state nucleon.
The CGLN amplitude $F_i$ is the function of energy and angle
$\theta$ given by
\begin{eqnarray}\label{nr-2}
F_1&=&C_+\left[\,{\cal
A}_1+\frac{q\cdot k}{W-M}\,({\cal A}_3-{\cal A}_4)+(W-M')\,{\cal
A}_4\right]\,,\nonumber\\%
F_2&=&C_-\left[\,{\cal A}_1-\frac{q\cdot k}{W+M}\,({\cal
A}_3-{\cal A}_4) -(W+M')\,{\cal
A}_4\,\right]\,,\nonumber\\%
F_3&=&D_+\left[\,(W-M)\,{\cal A}_2+({\cal A}_3-{\cal A}_4)
\,\right]\,,\nonumber\\%
F_4&=&D_- \left[\,-(W+M)\,{\cal A}_2+({\cal A}_3-{\cal
A}_4)\,\right]\,,
\end{eqnarray}
with the normalization constants
$C_\pm=\frac{|\vec{k}|}{4\pi}\sqrt{\frac{E'\pm M'}{2W}}$ and
$D_\pm=\frac{|\vec{k}||\vec{q}|}{4\pi W}\sqrt{\frac{E'\pm
M'}{2W}}\ $. The $E(E')$ is the initial(final) state nucleon
energy. The invariant amplitude ${\cal A}_i$ in Eq. (\ref{nr-2})
is given by the following decomposition of the photoproduction
amplitude,
\begin{eqnarray}\label{m1}
{\cal M}=\bar{u}^{\prime}(p')\sum_{i=1}^{4}\,\gamma_5\,{\cal
A}_i\,M_{i}\,u(p)
\end{eqnarray}
where
\begin{eqnarray}\label{op}
&&M_1=\frac{1}{2}(\,/\kern-6pt{\epsilon}\,/\kern-6pt{k}-/\kern-6pt{k}
\,/\kern-6pt{\epsilon}\,)\,,\nonumber\\%
&&M_2=2P\cdot k\, q\cdot\epsilon-2P\cdot\epsilon\, q\cdot
k\,,\nonumber\\%
&&M_3=q\cdot
k\,/\kern-6pt{\epsilon}-q\cdot\epsilon\,/\kern-6pt{k}\,,\nonumber\\%
&&M_4=2(P\cdot
k\,/\kern-6pt{\epsilon}-P\cdot\epsilon\,/\kern-6pt{k})-(M+M')M_1\,,
\end{eqnarray}
are the transition operators with $P=\frac{1}{2}(p+p')$. We use
the conventions of Bjorken and Drell \cite{bjorken} through out
this work. Note that we adopt the covariant operators, Eq.
(\ref{op}), in the decomposition of the amplitude in order for the
$t$-channel pion pole to be free of the kinematic singularities
\cite{dennery}. (In the case of electroproduction process the
decomposition used in Ref. \cite{thom} does not guarantee this
condition.)

Therefore, the Regge pole exchange in the $t$-channel is
incorporated in the SCHA $H_i$ through the reggeization of the
fixed-$t$ pole in the amplitude ${\cal M}$, which is usually given
by the conventional Born terms. The cross section and spin
polarization observables formulated in terms of the $H_i$ are
presented in Table \ref{tb1}. For comparison the definitions of
the helicity amplitude used by other authors are collected as
well. In the Regge regime, $s\gg 4M^2$ and the small $-t$, the
general features of the two different helicity formulations, the
TCHA, and the SCHA are discussed in Appendix  A.
\begin{table*}{}
\caption{\label{tb1} Notations for the $s$-channel helicity
amplitudes. (a)\cite{npb95}, (b)\cite{npb37}, and
(c)\cite{walker}. The net helicity flip is denoted by $n$. The
variable $t=t_{min}-4|\vec{k}||\vec{q}|\sin^2\frac{\theta}{2}$\,,
where $t_{min}=(k_0-q_0)^2-(|\vec{k}|-|\vec{q}|)^2$.}
\begin{ruledtabular}\label{tb1}
\begin{tabular}{cccccl}
                  & Baker$^{(a)}$& Worden$^{(b)}$ & Walker$^{(c)}$ &  Observables of Walker      &\\%
\hline
$n=0$              &$N$    &  $H_2$   & $H_2$   & $d\sigma/dt=|H_1|^2+|H_2|^2+|H_3|^2+|H_4|^2$   &\\%
$n=1$              &$S_1$  & $H_4$    & $H_1$  & $\Sigma d\sigma/dt=2{\rm Re}(H_1^*H_4-H_2^*H_3)$\\%
$n=1$              &$S_2$  & $H_1$    & $H_4$  & $T d\sigma/dt=2{\rm Im}(H_1^*H_2-H_3^*H_4)$     &\\%
$n=2$              & $D$  & $H_3$    & $H_3$   & $P d\sigma/dt=2{\rm Im}(H_2^*H_4-H_1^*H_3)$     &\\%
\hline
Normalization      &1      &$\sqrt{32\pi}(s-M^2)$&$\frac{s-M^2}{\sqrt{2\pi s}}$ &                &\\%
\end{tabular}
\end{ruledtabular}
\end{table*}

To express the four channels for charged and neutral pion
photoproductions, the invariant amplitude is decomposed into the
following form in isospin space:
\begin{eqnarray}
{\cal A}_i={\cal A}_i^{(+)}\delta_{a3}+{\cal
A}_i^{(-)}\frac{1}{2}[\tau_a,\,\tau_3]+{\cal A}_i^{(0)}\tau_a\,,
\end{eqnarray}
and the four respective amplitudes for particular physical
processes are given by
\begin{eqnarray}\label{iso}
&&{\cal A}_i(\gamma p\to \pi^+ n)=\sqrt{2}({\cal A}^{(0)}_i+{\cal
A}^{(-)}_i)\,,\nonumber\\%
&&{\cal A}_i(\gamma n\to \pi^-
p)=\sqrt{2}({\cal A}^{(0)}_i-{\cal A}^{(-)}_i)\,,\nonumber\\%
&&{\cal A}_i(\gamma p\to \pi^0 p)={\cal A}^{(0)}_i+{\cal
A}^{(+)}_i\,,\nonumber\\%
&&{\cal A}_i(\gamma n\to \pi^0 n)=-{\cal A}_i^{(0)}+{\cal
A}^{(+)}_i\,.
\end{eqnarray}

\subsection{Charged pion photoproduction}

\begin{figure}[htbp]
\includegraphics*[width=8.6cm]{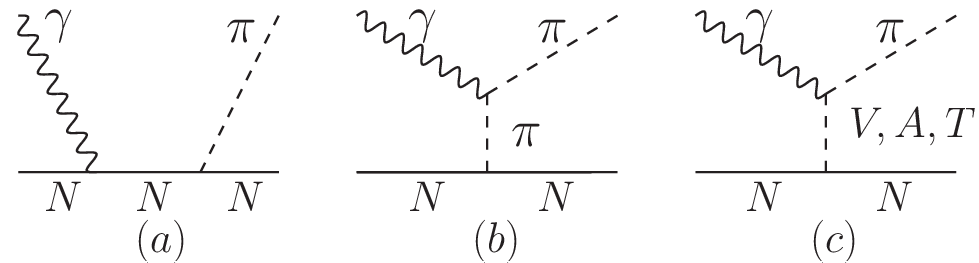}
\caption[]{(Color online) Exchange of $t$-channel mesons. Diagram
(a) is the nucleon Born term and (b) is the pion Regge pole
exchange. The diagrams (a) and (b) constitute the gauge invariant
pion Regge pole exchange. Diagram (c) represents the vector
meson(V), axial vector meson(A), and tensor meson(T) Regge pole
exchanges. } \label{fig:diag02}
\end{figure}

We now consider the reggeization of the $t$-channel meson pole on
the basis of the Born approximation to the first order
approximation of one photon exchange \cite{cgln}. Following Ref.
\cite{guid} we introduce the nucleon Born term to restore gauge
invariance of the $\pi$N system coupling to photon at the
tree-level Feynman diagram. We then make a prescription for the
reggeization of the $t$-channel pion exchange by replacing the
fixed-$t$ pole with the Regge propagator, while keeping intact the
coupling vertices $\gamma\pi\pi$ and $\pi$NN given by the
effective Lagrangians. Therefore, the reggeized pion pole
exchanges are written as
\begin{widetext}
\begin{eqnarray}\label{mpi+}
&&{\cal M}_{\pi^+ }=i\sqrt{2}\,eg_{\pi NN}\,\bar{u}'(p^{\prime})
\left[\gamma_5\frac{(2q-k)\cdot\epsilon}{t-m^2}+\gamma_5\frac{/\kern-6pt
p + /\kern-6pt k + M}{s-M^2} \left(/\kern-6pt\epsilon
-\frac{\kappa_p}{2M}\,/\kern-6pt{\epsilon}\,/\kern-6pt{k}\,\right)
\right](t-m^2)
{\cal P}^\pi(s,t)\,u(p),\\%
&&{\cal M}_{\pi^-}=-i\sqrt{2}\,eg_{\pi NN}\,\bar{u}'(p^{\prime})
\left[\gamma_5\frac{(2q-k)\cdot\epsilon}{t-m^2}-\left(/\kern-6pt\epsilon
-\frac{\kappa_p}{2M}\,/\kern-6pt{\epsilon}\,/\kern-6pt{k}\,\right)
\frac{/\kern-6pt{p'}- /\kern-6pt k + M'}{u-{M'}^2} \gamma_5
\right](t-m^2){\cal P}^\pi(s,t)\,u(p), \label{mpi-}
\end{eqnarray}
\end{widetext}
which are gauge invariant for the $\gamma p\to \pi^+n$ and $\gamma
n\to \pi^-p$ processes. Here the $g_{\pi NN}$ is the $\pi$N strong
coupling constant, $m$ is the pion mass, and
\begin{eqnarray}\label{pi-regge}
{\cal
P}^\pi(s,t)=\frac{\pi\alpha'_\pi}{\Gamma(\alpha_\pi(t)+1)}\frac{(1+
e^{-i\pi\alpha_\pi(t)})}{2\sin\pi\alpha_\pi(t)}\left(\frac{s}{s_0}\right)^{\alpha_\pi(t)}
\end{eqnarray}
is the pion Regge propagator,
which leads Eqs. (\ref{mpi+}) and (\ref{mpi-}) to the usual
nucleon and pion Born terms in the limit $t\approx m_\pi^2$, as
\begin{eqnarray}\label{lim}
(t-m^2){\cal P}^\pi(s,t)\to 1\ .
\end{eqnarray}
This on-mass shell relation between the two propagators is easily
proved by using the properties of the $\Gamma$ function
\cite{sibi}. In Fig. \ref{fig:diag02} the diagrams (a) and (b)
show the gauge invariant nucleon and pion Born terms in Eqs.
(\ref{mpi+}) and (\ref{mpi-}) reggeized through the prescription
in Eq. (\ref{lim}). At high energy it is sufficient to neglect the
magnetic interaction of the nucleon Born term. In this regard it
would be redundant to consider the pseudovector coupling of the
$\pi$N interaction which differs from the pseudoscalar coupling
one only by the magnetic interaction of the Seagull term.

The photoproduction amplitudes for the spin-1 vector meson, and
axial vector meson exchange are given by
\begin{widetext}
\begin{eqnarray}\label{vec}
&&{\cal M}_{V}=\frac{g_{\gamma \pi V}}{m_0}
\epsilon_{\mu\nu\alpha\beta}\epsilon^{\mu}k^{\nu}q^{\alpha}
\frac{(-g^{\beta\rho}+
Q^{\beta}Q^{\rho}/m_{V}^2)}{t-m_{V}^2+im_{V}\Gamma_{V}}
\bar{u}'(p')\left[ g^v_{VNN}\,\gamma_{\rho} +
i\frac{g^t_{VNN}}{2M}\sigma_{\lambda\rho}Q^{\lambda} \right]u(p)\
,\\%
&&{\cal M}_{A}=i\frac{g_{\gamma \pi A}}{m_0} ( k\cdot
Q\,\epsilon_{\mu} -\epsilon\cdot Q\,k_{\mu})
\frac{(-g^{\mu\nu}+Q^{\mu}Q^{\nu}/m_A^2)}{t-m_A^2
+im_A\Gamma_A}\bar{u}'(p')\left[g^v_{ANN}\,\gamma_{\nu}+
i\frac{g^t_{ANN}}{2M} \sigma_{\lambda\nu}Q^{\lambda}
\right]\gamma_5 u(p)\ , \label{ax}
\end{eqnarray}
\end{widetext}
respectively, where $m_0$ is a parameter of mass dimension taken
as $1$ GeV and $Q=(q-k)$ is the $t$-channel momentum transfer
\cite{bgyu1}.

The exchange of the tensor meson of spin-2 is given by
\cite{giac,ysoh,klei}
\begin{widetext}
\begin{eqnarray}\label{tensor}
{\cal M}_{T}
&=&\bar{u}'(p')\,\varepsilon_{\alpha\beta\mu\nu}\,\epsilon^\mu\,k^\nu
q^\alpha
q_\rho\frac{\Pi^{\beta\rho;\lambda\sigma}(q-k)}{t-m^2_T+im_T\Gamma_T}
[\,G^{(1)}_{T}(\gamma_\lambda P_\sigma+\gamma_\sigma
P_\lambda)+G^{(2)}_{T} P_\lambda P_\sigma\,]u(p)\,,
\end{eqnarray}
\end{widetext}
where the tensor meson coupling constants are
\begin{eqnarray}
G^{(1)}_{T}=\frac{2g_{\gamma \pi
T}}{m^2_0}\frac{2g^{(1)}_{TNN}}{M}\ , \ \
G^{(2)}_T=-\frac{2g_{\gamma \pi
T}}{m^2_0}\frac{4g^{(2)}_{TNN}}{M^2}\ ,
\end{eqnarray}
for brevity and the polarization tensor of the tensor meson is
\begin{eqnarray}
&&\Pi_{\mu\nu;\rho\sigma}(Q)=\frac{1}{2}(\bar{g}_{\mu\rho}\bar{g}_{\nu\sigma}
+\bar{g}_{\mu\sigma}\bar{g}_{\nu\rho})-\frac{1}{3}\bar{g}_{\mu\nu}\bar{g}_{\rho\sigma}\,,
\nonumber\\%
&&(\,\bar{g}_{\mu\nu}=-g^{\mu\nu} + Q^{\mu}Q^{\nu}/m_T^2)\,.
\end{eqnarray}
These $t$-channel meson exchanges are depicted in Fig.
\ref{fig:diag02}(c). In the above expressions we write
collectively the vector meson, $V$=$\rho(770)$($1^+(1^{--})$), and
$\omega(782)$($0^-(1^{--})$) with the isospin and spin quantum
number denoted by $I^G(J^{PC})$. For the axial meson,
$A$=$a_1(1260)$($1^-(1^{++})$), $b_1(1235)$($1^+(1^{+-})$), and
$h_1(1170)$($0^-(1^{+-})$). The tensor meson
$a_2(1320)$($1^-(2^{++})$) is denoted by the $T$.

With the Regge propagator for the pion exchange given in Eq.
(\ref{pi-regge}), the reggeization of the vector meson, axial
meson, and the tensor meson exchanges in the photoproduction
amplitude follows by replacing each fixed-$t$ pole in Eqs.
(\ref{vec}), (\ref{ax}) and (\ref{tensor}) with the corresponding
Regge propagator of the form
\begin{eqnarray}\label{vec-regge}
&&
{\cal
P}^{V}(s,t)=\frac{\pi\alpha'_V}{\Gamma(\alpha_V(t))}\frac{(-1+
e^{-i\pi\alpha_V(t)})}{2\sin\pi\alpha_V}\left(\frac{s}{s_0}\right)^{\alpha_V(t)-1}\,,\nonumber\\
&&
{\cal P}^A(s,t)=\frac{\pi\alpha'_A}{\Gamma(\alpha_A(t))}\frac{(-1+
e^{-i\pi\alpha_A(t)})}{2\sin\pi\alpha_A}\left(\frac{s}{s_0}\right)^{\alpha_A(t)-1}\,,\nonumber\\
&&
{\cal
P}^{T}(s,t)=\frac{\pi\alpha'_T}{\Gamma(\alpha_T(t)-1)}\frac{(1+
e^{-i\pi\alpha_T(t)})}{2\sin\pi\alpha_T(t)}\left(\frac{s}{s_0}\right)^{\alpha_T(t)-2}\,,
\end{eqnarray}
respectively. The scale parameter $s_0$ is taken as $1$ GeV$^2$
for the variable $(s/s_0)$ to be dimensionless. The exchange non-degenerate (EXnD) phase factor
$\frac{1}{2}(\tau+\,e^{-i\pi\,\alpha(t)})$ with the signature
$\tau=(-1)^J$ in Eq. (\ref{vec-regge}) is concerned with the
notion of the exchange degeneracy (EXD) of $\rho$-$a_2$, or of
$\pi$-$b_1$ pair, as discussed in Appendix B.

As to the Regge trajectories of the form,
$\alpha(t)=\alpha'\,t+\alpha_0$, in Eqs. (\ref{pi-regge}) and
(\ref{vec-regge}), we have to allow somewhat uncertainties in the
choice of the slope $\alpha'$ given in units of GeV$^{-2}$ and the
intersection $\alpha_0$ \cite{kell,sibi,mosel,aniso}. In this work
we use the same ones chosen in Ref. \cite{guid} for the primary
meson exchanges $\pi$, $\rho$ and $\omega$. We take the
trajectories of the $b_1$ and $a_2$ to be the weakly EXD with
$\pi$ and $\rho$, respectively. The trajectories of the $a_1$ and
$h_1$ are chosen to be the same as that of $b_1$ with their own
mass and spin eigenstates specified.
\begin{eqnarray}\label{pitraj}
&&\alpha_\pi(t)=0.7\,(t-m^2_\pi)\,,\nonumber\\
&&\alpha_{\rho}(t)=0.8\,t+0.55\,,\nonumber\\
&&\alpha_{\omega}(t)=0.9\,t+0.44\,,\nonumber\\
&&\alpha_{a_1}(t)=0.7\,(t-m_{a_1}^2)+1\,,\nonumber\\
&&\alpha_{b_1}(t)=0.7\,(t-m_{b_1}^2)+1\,,\nonumber\\
&&\alpha_{h_1}(t)=0.7\,(t-m^2_{h_1})+1\,,\nonumber\\
&&\alpha_{a_2}(t)=0.8\,(t-m_{a_2}^2)+2\,.
\end{eqnarray}

For the charged pion case the vector meson $\omega$, $\phi$
exchanges and the Pomeron exchange are forbidden by charge
conservation.
The amplitudes for the
vector meson, the axial meson, and the tensor meson are given by
Eqs. (\ref{vec}), (\ref{ax}), and (\ref{tensor}), respectively.
The sign of each term follows from the consideration of $G$-parity
at the radiative decay vertex.
Note that $\gamma\pi \pi$, $\gamma\pi a_1$, and $\gamma\pi a_2$ change
sign, while the signs of $\gamma\pi b_1$ and $\gamma\pi \rho$ are not changed.
We now have two options on the
condition of each EXD pair, which is either strong or weak. The
former condition has been adopted in Refs.  \cite{levy,guid} to
exclude the exchange of $b_1$ and $a_2$. We here opt to choose the
weak EXD pair in which case the exchange of the Regge pole in the
pair has the coupling vertex different from each other but shares
a common phase. Therefore, we take the phases of EXD pairs,
$\pi$-$b_1$ and $\rho$-$a_2$ to be rotating
$e^{-i\pi\alpha(t)}$ for the $\gamma p\to\pi^+n$ process, whereas
the phases of these meson exchanges are chosen to be constant
for the process $\gamma n\to\pi^-p$, as explained in Appendix
B.

In the schematic notation the production amplitudes with the phases
of Regge poles for charged channels are written as
\begin{eqnarray}
\label{born+}
&&{\cal M}_{\pi^+ n}={\cal
M}_{\pi^+}\cdot e^{-i\pi\alpha_\pi}
+\sqrt{2}\,[{\cal M}_{b_1}\cdot e^{-i\pi\alpha_{b_1}}
\nonumber\\&&
+{\cal
M}_{\rho}\cdot (-e^{-i\pi\alpha_\rho})+{\cal
M}_{a_2}\cdot (-e^{-i\pi\alpha_{a_2}})+{\cal M}_{a_1}\cdot{\rm EXnD}],
\nonumber\\
\end{eqnarray}
for the $\gamma p\to\pi^+n$ process and
\begin{eqnarray}\label{born-}
&&{\cal M}_{\pi^- p} ={\cal M}_{\pi^-}\cdot 1+\sqrt{2}\,[{\cal
M}_{b_1}\cdot(-1)
+{\cal M}_{\rho}\cdot 1-{\cal M}_{a_2}\cdot 1
\nonumber\\&&
-{\cal M}_{a_1}\cdot{\rm EXnD}],
\end{eqnarray}
for the $\gamma n\to\pi^-p$ process, respectively. The $\pi^\pm$ terms are given
in Eqs. (\ref{mpi+}) and (\ref{mpi-}).
Axial mesons $a_1$ and $b_1$ refer to the vector and tensor part of
Eq. (\ref{ax}), respectively where the phase of the exchange non-degenerate (EXnD)
$a_1$ meson is chosen as in Eq. (\ref{vec-regge}).

%

\subsection{Neutral pion photoproduction}

The neutral pion photoproduction excludes the nucleon Born term
because the $t$-channel pion exchange is absent from the process
by charge conservation. This process is known to allow the one
photon exchange in the $t$-channel through
the $\pi^0\to \gamma\gamma$ decay known as the Primakoff effect
\cite{donn,laget}. It could play a role at the very forward angle
$-t\approx 0$. Excluding the Primakoff region, we find that
neither the two vacuum trajectories $P$ and $P'$ of even  parity
nor the trajectories associated with the axial meson $a_1$ and
tensor mesons $a_2$ and $f_2$ are allowed to decay to
$\pi^0\gamma$ by conservation of charge conjugation($C$).
Therefore, the exchange of the Regge poles for neutral pion
photoproduction is composed of the vector mesons $\omega$ and
$\rho$, and the axial meson $b_1$. In addition, we note that the
axial mesons $h_1(1170)$ and $h_1(1380)$, the singlet and octet
members of the axial meson nonet to which the $b_1$ belongs, can
also contribute. In consideration of Ref.  \cite{roca} where the
radiative decay width $\gamma\pi h_1(1170)$ was predicted from the
coupled channel analysis we find it to be comparable to that of
$b_1$. We include the axial meson $h_1(1170)$ exchange to
contribute as the unnatural parity exchange together with the
$b_1$.

From the isospin relation in Eq. (\ref{iso}) we have
\begin{eqnarray}\label{born0}
&&{\cal M}_{\pi^0 p}={\cal M}_{\omega}+{\cal M}_{\rho}
+{\cal M}_{b_1}+{\cal M}_{h_1} \,,\\
&&{\cal M}_{\pi^0 n}={\cal
M}_{\omega}-{\cal M}_{\rho}-{\cal M}_{b_1}
+{\cal M}_{h_1}\,, \label{born01}
%
\end{eqnarray}
for the processes $\gamma p\to\pi^0
p$ and $\gamma n\to\pi^0 n$, respectively.

Since the Regge poles
in Eqs. (\ref{born0}) and (\ref{born01}) are not EXD with each
other, we take the non-degenerate phase,
$\frac{1}{2}(-1+e^{-i\pi\alpha_\omega})$, for the leading $\omega$
trajectory and, in principle, do the same for the rest in both
processes. However, we may as well have more freedom to choose the
rotating phase $e^{-i\pi\alpha_\rho}$ for the $\rho$ and the
constant one for the axial meson $b_1$ and $h_1$ in Eq.
(\ref{born0}). In Eq. (\ref{born01}) we also choose the phase of
the $\omega$ to be non-degenerate, but we take the constant phase
for the rest for a better description of the phenomenology as we
shall show later.

Before closing this section, it is worth remarking that the
discontinuity of the cross section such as a sharp rise or an
apparent dip might require a consideration of the Regge cuts
arising from the multiple interferences between two or more Regge
poles \cite{kell,sibi,heny,laget1,laget2}. In this work, however,
we disregard such higher order effects mainly because we avoid
introducing unwanted parameters through the cut. It is, therefore,
sufficient to check the consistency of our amplitudes given in
Eqs. (\ref{born+}), (\ref{born-}), (\ref{born0}), and
(\ref{born01}) with high-energy data in the region, $-t\leq 2 $
GeV$^2$ and $s\gg t$.

\section{Determination of meson coupling constants}

This section is devoted to the determination of the coupling
constants of exchanged mesons by using the axiomatic identities
which are based on the vector meson dominance (VMD) for the vector
meson and the axial vector meson dominance (AVMD) for axial vector
meson coupling to the nucleon, respectively. The validity of the
tensor meson dominance (TMD) is discussed for the determination of
the tensor meson-nucleon coupling constant. These meson-baryon
coupling constants are determined by basically respecting the
SU(3) relations. A consistency check of the coupling constants
follows by comparing the results with existing estimates that are
widely accepted in other reaction processes.

\subsection{Vector meson couplings}

For the determination of $\rho$NN coupling constants, we refer to
the VMD which assumes the dominance of the $\rho^0$ in the nucleon
electromagnetic form factors coupling to photon \cite{sakurai},
i.e.,
\begin{eqnarray}\label{vmd00}
<N|j^\mu(0)|N>=\frac{<0|j^\mu(0)|\rho><\rho,N|N>}{t-m_{\rho}^2}\,,
\end{eqnarray}
where the nucleon isovector form factors are defined as
\begin{eqnarray}\label{rho-ff}
<N|j_a^\mu(0)|N>=\bar{u}(p')[F_1(t)\gamma^\mu+F_2(t)i\sigma^{\mu\nu}
q_\nu]\frac{\tau_a}{2}u(p)\,,
\end{eqnarray}
and $t$ is the four-momentum transfer $q=(p'-p)$ squared. The
tensor coupling part vanishes at $q=0$. The $\rho$ meson decay is
given by the current-field identity \cite{sakurai1},
\begin{eqnarray}\label{rho-decay}
<0|j^\mu(0)|\rho>= \frac{m_\rho^2}{f_{\rho}}\epsilon^\mu\ .
\end{eqnarray}
The $m_\rho$ and $\epsilon^\mu$ are the $\rho$ meson mass and the
polarization vector, respectively. The $f_\rho$ is the universal
coupling constant of the $\rho$ meson which is determined by the
decay $\rho^0\to e^+e^-$.
Therefore, with the $\rho NN$ coupling given by
\begin{eqnarray}
<\rho,N|N>=g^v_{\rho NN}\,\bar{u}(p')\gamma^\nu\tau_a\,
u(p)\epsilon_\nu\,,
\end{eqnarray}
the VMD in Eq. (\ref{vmd00}) yields the $t$-dependence of the
isovector form factor
\begin{eqnarray}\label{vmd01}
\frac{1}{2}F_1(t)=\frac{m_\rho^2}{f_\rho}\frac{1}{m_\rho^2-t}g_{\rho
NN}\,,
\end{eqnarray}
which requires at $t=0$
\begin{eqnarray}
g^v_{\rho NN}=\frac{1}{2}f_\rho\ .
\end{eqnarray}
The value for the $f_\rho$ varies in the range from the
$f_\rho=4.94$ estimated from the decay width $\Gamma_{\rho^0\to
e^+e^-}=7.04$ keV to the case of $f_{\rho\pi\pi}=6.01$ obtained by
the width $\Gamma_{\rho^0\to\pi^+\pi^-}=149.4$ MeV. On the other
hand, the VMD applied to the low energy $s$-wave $\pi$N scattering
leads to $\sqrt{f_{\rho\pi\pi}f_{\rho NN}}=5.85$.  Also there
exits an estimate $f_\rho\approx 5.3$ from the empirical analysis
of the experimental data on the $N\bar{N}\pi\pi$ \cite{rho,npb95}.
In this work we choose the $f_\rho=5.2$, rather a moderate one
from Refs. \cite{rho,npb95}, and use
\begin{eqnarray}
g_{\rho NN}=2.6\ .
\end{eqnarray}
For the ratio of tensor to vector coupling, $\kappa_\rho$, the VMD
leads to $\kappa_\rho=3.7$, whereas $\kappa_\rho\approx 6$ is
extracted from the one boson exchange (OBE) analysis of the NN
potential \cite{weise}. We use $\kappa_\rho=6.2$ for the present
work.
For the $\omega NN$ coupling constants, we take $g_{\omega
NN}=15.6$ by the ratio $g_\omega=f_\omega=3f_\rho$ with the tensor
coupling ratio $\kappa_\omega\simeq 0$.

The coupling constant of the $\gamma\pi V$ interaction is
estimated from the observed decay width. From the effective
Lagrangian used in Eq. (\ref{vec}),
\begin{eqnarray}
{\cal L}_{\gamma\pi V}=\frac{g_{\gamma\pi
V}}{2m_0}\widetilde{F}_{\mu\nu}\,V^{\mu\nu}\pi \,\,,
%
\end{eqnarray}
with photon pseudotensor field $\widetilde{F}_{\mu\nu}={1\over2}\epsilon^{\mu\nu\alpha\beta}F_{\alpha\beta}$,
the decay width is given by,
\begin{eqnarray}\label{vec-decay}
\Gamma_{V\to \pi\gamma}=\frac{1}{96\pi}\left(\frac{g_{\gamma\pi V
}}{m_0}\right)^2 \left(\frac{m_{V}^2-m^2_\pi}{m_{V}}\right)^3\ ,
\end{eqnarray}
which estimates $g_{\gamma\pi^\pm\rho}=0.223$ from the measured
width $\Gamma_{\rho\to\gamma\pi^\pm}=0.068$ GeV and
$g_{\gamma\pi^0\rho}=0.255$ from the
$\Gamma_{\rho\to\gamma\pi^0}=0.090$ GeV, respectively. These
coupling constants are in fair agreement with the prediction of
the VMD in the $\omega\rho\pi$ coupling vertex, which states in
the manner similar to Eq. (\ref{vmd01}),
\begin{eqnarray}\label{vmd1} g_{\gamma\pi
\rho}=e\frac{m^2_\omega}{f_\omega}\frac{1}{m^2_\omega-t}\,g_{\omega\rho\pi}
\end{eqnarray}
with $g_{\omega\rho\pi}=\frac{3f_\rho^2}{8\pi^2 f_\pi}=11.035$
GeV$^{-1}$ \cite{anomaly} estimated by $f_\rho=5.2$,
$f_\omega=3f_\rho$, and the pion decay constant $f_\pi=93.1$ MeV.

For $\gamma\pi^0\omega$ coupling, we determine
$g_{\gamma\pi^0\omega}=0.723$ from the empirical decay width
$\Gamma_{\omega\to\gamma\pi^0}=0.757$ GeV, which is comparable to
the VMD prediction similar to Eq. (\ref{vmd1}),
\begin{eqnarray}
g_{\gamma\pi^0\omega}= \frac{e}{f_\rho}g_{\omega\rho\pi}\ .
\end{eqnarray}

\subsection{Axial meson couplings}

Radiative decays of axial mesons, $A\to \gamma\pi$, have been
investigated in Refs.  \cite{xio,hagl}, in which the coupling
constants of the mesons were derived from the VMD via the
interpolation of the $\rho(\omega)$ meson field into the strong
decay $A\to \rho\pi\,(A\to \omega\pi)$ for the charged (neutral)
meson couplings. Compared with the empirically known cases,
predictions by the VMD seem to be reliable, as shown above.
Hence we exploit the $\omega$ dominance in the decay
$b_1\to\omega\pi^0$ for the estimate of the $\gamma\pi^0 b_1$
coupling constant the width of which is currently not known.
By using the effective Lagrangian for the photon-pseudoscalar
meson($\pi$)-axial meson($A$) coupling,
\begin{eqnarray}\label{eff1}
{\cal L}_{\gamma\pi A}=\frac{g_{\gamma\pi
A}}{2m_0}F_{\mu\nu}A^{\mu\nu}\,\pi\,,
%
\end{eqnarray}
we derive the decay width as in Ref. \cite{hagl} which
estimates $g_{b_1\omega\pi^0}=9.77$ from the full width
$\Gamma_{b_1\to\pi\omega}=(142\pm 9)$ MeV. Thus, we obtain
$g_{\gamma\pi^0 b_1}=0.189$ from the VMD relation,
\begin{eqnarray}\label{ax0}
g_{\gamma\pi^0 b_1}=\frac{e}{f_\omega}g_{b_1 \omega\pi^0}\, .
\end{eqnarray}

For the $b_1$ decay into the charged pion case the width is
reported in the Particle Data Group to be $\Gamma_{b_1\to
\pi^\pm\gamma}=(0.227\pm 0.057)$ MeV, and we estimate the coupling
constant to be $g_{\gamma\pi^\pm b_1}=0.196$ by use of Eq.
(\ref{vec-decay}). These values are comparable with the
chiral-unitary model predictions, $g_{\gamma\pi^\pm b_1}=0.187$,
and $g_{\gamma\pi^0 b_1}=0.173$ \cite{roca}. The empirical
information on the decays of the isoscalar $h_1(1170)$ and
$h_1(1380)$ mesons is very scarce. We refer to the chiral-unitary
model predictions for the decay widths
$\Gamma_{h_1(1170)\to\pi^0\gamma}=837\pm 134$ keV and
$\Gamma_{h_1(1380)\to\pi^0\gamma}=81\pm 18$ keV, which yield
$g_{\gamma\pi^0 h_1(1170)}=0.405$ and $g_{\gamma\pi^0
h_1(1380)}=0.098$, respectively. Thus the $h_1(1380)$ exchange is
neglected for the small coupling constant hereafter.

The case of determining the $g_{\gamma\pi a_1}$ coupling constant
is somewhat uncertain, because only the full width is given in a
broad range, $\Gamma=250\sim 600$ MeV. Therefore, even if we use
the VMD to estimate the coupling constant, as before,
\begin{eqnarray}\label{ax1}
g_{\gamma\pi^\pm a_1}=\frac{e}{f_\rho}g_{\rho\pi^\pm a_1}\,,
\end{eqnarray}
we still need to know the partial decay width
$\Gamma_{a_1\to\pi\rho}$ for the determination of the
$g_{\rho\pi^\pm a_1}$. For this reason Xiong \cite{xio} and Haglin
\cite{hagl} assumed the width $\Gamma_{a_1\to\pi\rho}=400$ MeV to
obtain $g_{\gamma\pi^\pm a_1}=0.743$. We find, however, that this
value yields the partial width $\Gamma_{a_1\to\pi\gamma}=1.4$ MeV,
which certainly overestimates the experimental value $0.64\pm
0.246$ MeV \cite{amsler,ziel,pdg1}. In this work we choose the
$\Gamma_{a_1\to \gamma\pi}$ = 0.64 MeV to estimate
$g_{\gamma\pi^\pm a_1}=0.316$, which corresponds to the width
$\Gamma_{a_1\to\pi\rho}\approx 250$ MeV. This choice is
reasonable, as compared to the chiral unitary model estimate where
the decay width was predicted to be
$\Gamma_{a_1\to\pi\gamma}$=0.46$\pm$0.1 MeV, and $g_{\gamma\pi
a_1}=0.268$, as a result \cite{roca}. We note that this value is
within the range of the widths
0.630$\pm$0.246 MeV.

For the determination of the axial meson coupling constants,
$g^v_{ANN}$, and $g^t_{ANN}$ in Eq. (\ref{ax}), we apply the AVMD
to the nucleon axial form factors \cite{bir,gam}
\begin{eqnarray}\label{ax-ff}
<N|j_{5\,a}^\mu(0)|N>
=\bar{u}(p')[F_A(t)\gamma^\mu +F_T(t)i\sigma^{\mu\nu}q_\nu
]\gamma_5\frac{\tau_a}{2}u(p),
\end{eqnarray}
where $F_A(t)$ is the axial vector form factor with $F_A(0)=g_A$,
and $F_T(t)$ is the pseudotensor one. The induced pseudoscalar
form factor $F_P(t)q^\mu\gamma_5$ is omitted here for irrelevance.
Since the nucleon axial vector vertex $\gamma^\mu\gamma_5$ is
$C$-even and the pseudotensor coupling vertex
$\sigma^{\mu\nu}\gamma_5$ is $C$-odd, the $a_1$ meson of $C$-even
couples to the nucleon via the axial vector coupling only, whereas
both the $b_1$ and $h_1$ mesons of $C$-odd couple to the nucleon
axial current with the pseudotensor coupling, but not with the
axial vector coupling \cite{mosel}.

\subsubsection{$a_1$ vector coupling constant}

In analogy to the $\rho$ meson dominance as before, the nucleon
axial vector current is assumed to be dominated by the $a_1$-pole,
and this idea is known to hold up to the momentum transfer
$t\approx$ 1 GeV$^2$ for the optimal fit of the axial form factor
$F_A(t)$ to experimental data \cite{gari}. Thus, the matrix
element of the nucleon axial vector current is
\begin{eqnarray}\label{avmd00}
<N|j_5^\mu(0)|N>=\frac{<0|j^\mu_5(0)|a_1><a_1,N|N>}{t-m_{a_1}^2}\,.
\end{eqnarray}
The charged $a_1$ meson decay to vacuum through the axial vector
current is given by
\begin{eqnarray}\label{ax-decay}
<0|j^\mu_5(0)|a_1>= \frac{f_{a_1}}{\sqrt{2}}\epsilon^\mu\,,
\end{eqnarray}
where the $a_1$ decay constant $f_{a_1}=(0.19\pm0.03)$ GeV$^2$ is
measured from the $\tau^-\to a_1^-+\nu_\tau$ decay process. The
$a_1$ coupling to nucleon is given by
\begin{eqnarray}
<a_1,N|N>=g^v_{a_1NN}\,\bar{u}(p')\gamma^\nu\gamma_5\,\tau_a\,
u(p)\epsilon_\nu.
\end{eqnarray}
Then, we obtain the relation between the axial vector form factor
and the strong coupling vertex of the axial meson at $t=0$ in Eq.
(\ref{avmd00}) \cite{bir},
\begin{eqnarray}\label{avmd}
g^v_{a_1NN}=\frac{1}{2}\frac{\sqrt{2}\,m_{a_1}^2}{f_{a_1}}g_A\,.
\end{eqnarray}
We are now able to determine $g^v_{a_1NN}$ with the nucleon axial
charge $g_A=1.25$.
But the EMC experiment reported rather a smaller value,
\begin{eqnarray}
g_A=(\Delta u-\Delta d)=1.134\,,
\end{eqnarray}
measured in terms of the quark helicity $\Delta q$ involving the
sea quarks and gluon contributions. This leads to
\begin{eqnarray}
g^v_{a_1NN}=6.7\,,
\end{eqnarray}
and we find it to be in agreement with the $g^2_{a_1 NN}/4\pi=3.3$
extracted from the analysis of NN interaction \cite{durso}.

\subsubsection{$b_1$ $\&$ $h_1$ tensor coupling constant}

In order to determine the tensor coupling constants of the $b_1NN$
and $h_1NN$ interactions, we apply the AVMD to the nucleon
pseudotensor form factor,
\begin{eqnarray}\label{ax-tensor}
<N|j_5^{\mu\nu}(0)|N>=\sum_{A=b_1,h_1}\frac{<0|j^{\mu\nu}_5(0)|A><A,N|N>}{m_{A}^2-t}\,,
\end{eqnarray}
with the pole dominance in Eq. (\ref{ax-tensor}) allowed for the
$b_1$ and $h_1$, as discussed. The axial meson decay via the
pseudotensor current is written as
\begin{eqnarray}\label{ax-tensor00}
<0|j^{\mu\nu}_5(0)|A>
=if_{A}(\epsilon^\mu q^\nu-\epsilon^\nu q^\mu),
\end{eqnarray}
with the axial meson decay constant $f_A$. The decay constants are
taken as $f_{b_1}=\frac{\sqrt{2}}{m_{b_1}}f_{a_1}\simeq 0.21$ GeV
and $f_{h_1}=f_{b_1}$ \cite{gam}. The strong coupling of the axial
meson to the nucleon pseudotensor current is given by
\begin{eqnarray}\label{100}
<A,N|N>=\frac{g^t_{ANN}}{2M}\bar{u}(p, s_T
)i\sigma^{\alpha\beta}\gamma_5\,q_\beta\epsilon_\alpha
\,\tau_a\,u(p, s_T),
\end{eqnarray}
and we identify the nucleon pseudotensor form factor in Eq.
(\ref{ax-tensor}) with the matrix element of the bilinear quark
tensor current \cite{gam},
\begin{eqnarray}\label{trans}
<N(p,s_T)|\bar{q}_a\,\sigma^{\mu\nu}\gamma_5\frac{\lambda_a}{2}q_a|N(p,s_T)>\nonumber\\
=2\,\delta q^a(\mu^2)(p^\mu s^\nu_T-p^\nu s^\mu_T)\,,
\end{eqnarray}
where
$J^{\mu\nu}_{5\,a}=\bar{q}_a\,\sigma^{\mu\nu}\gamma_5\frac{\lambda_a}{2}q_a$
and the $q_a$ denotes the quark field of favor $a=u,d, s$.
$N(p,s_T)$ is a nucleon state of momentum $p$ and spin
transversity $s_T$ \cite{jaffe}. The $\delta q^a$ is the quark
transversity which counts the valence quarks of opposite
transversities in the transversely polarized nucleon. The
measurement of the $\delta q^a$ depends on the renormalization
point $\mu^2$, since the chiral-odd pseudotensor current given in
Eq. (\ref{trans}) is not conserved. Thus, in contrast to the axial
charge, the nucleon tensor charge is determined by the quark
transversity $\delta q^a$ measured at the renormalization point
$\mu^2$ in Eq. (\ref{trans}).

Combining the above ingredients with each other in Eq.
(\ref{ax-tensor}), we obtain \cite{gam}
\begin{eqnarray}\label{tensorcc}
&&g^t_{b_1NN}=(\delta u-\delta
d)\frac{\sqrt{2}\,m_{b_1}^2\,M}{f_{b_1}<q^2_{\perp}>}\,,
\nonumber\\%
&&g^t_{h_1NN}=(\delta u+\delta
d)\frac{\sqrt{2}\,m_{h_1}^2\,M}{f_{h_1}<q^2_{\perp}>}\,,
\end{eqnarray}
where the $<q^2_{\perp}>$ is the quark transverse momentum squared
inside the nucleon. We, therefore, determine the tensor coupling
constant of the axial meson from the knowledge of the quark
transversity $\delta q^a$ and the quark transverse momentum
together with the decay constants given above. According to Ref.
\cite{gam} the quark transverse momentum squared in Eq.
(\ref{tensorcc}) is found to vary within the range,
$<q_\perp^2>\simeq$ 0.58 $\sim$ 1.0 GeV$^2$, inside the nucleon.
Hence, the present approach to the estimate of the $g_{ANN}^t$
must allow the uncertainty depending on the intrinsic quark
momentum as well as the renormalization point of the quark
transversity.
Furthermore, as the precise measurement of these latter quantities
are still in progress, the application of the AVMD to the nucleon
axial tensor form factor might be in question. However, as we
shall show in Table \ref{tb3} below, our estimate for the coupling
constant $g^t_{b_1NN}(g^t_{h_1NN})$ from Eq. (\ref{tensorcc}) is
valid within the range of the nucleon axial tensor charge
predicted by various QCD-inspired models.
%

As to the estimate of the $\delta q^a$ the model-independent
inequalities, though approximated, are known to be $|\delta
u|<3/2$ and $|\delta d|<1/3$ \cite{soffer}.
There also exist scattered values for the $\delta q^a$ from
various model calculations depending on what renormalization point
they adopted \cite{he, hckim,lat}. Table \ref{tb3} summarizes
existing estimates for the $\delta q^a$ and the $g^t_{ANN}$ as a
result. We choose the $\delta q^a$ from the light cone model
\cite{soff} in estimating the $g^t_{ANN}$ with signs in Table
\ref{tb4} favorable to reproduce the photoproduction data within
the range, because the quark helicities $\Delta u$ and $\Delta d$
from the model lead to the closest value for the nucleon axial
charge $g_A=1.25$ \cite{baron}.

It should be noted that our viewpoint in Eq. (\ref{tensorcc}) is
opposite to Ref. \cite{gam}; To be consistent with Eq. (\ref{ax})
we identify the interaction vertex in Eq. (\ref{100}) with the
tensor coupling constant $g_{ANN}^t$, and determine it by using
currently known values of the $\delta q^a$ based on the
QCD-inspired models. On the contrary the authors of Ref.
\cite{gam} used the known coupling constant $g_{ANN}$ in Eq.
(\ref{100})(which was determined from the SU(3) relation with the
$g_{a_1NN}^v$), and consequently in Eq. (\ref{tensorcc}), in order
to estimate the unknown $\delta q^a$.


%
\begin{table*}\caption{Tensor coupling constants of axial mesons from quark
transversities within the range $<q^2_{\perp}>=0.58\sim 1.0$
GeV$^2$. Model predictions for the quark transversity are quoted
from Ref. \cite{gam}. References are from the lattice \cite{lat},
QCD sum rule \cite{he}, MIT bag model \cite{bag}, constituent
quark model \cite{cqm}, quark soliton model \cite{hckim,soli1},
NJL model \cite{njl,njl1}, and light cone calculation \cite{soff}.
}
\begin{ruledtabular}\label{tb3}
\begin{tabular}{ccccccccl}
                  &Lattice & QCD sum rule  &  Bag & Constituent QM   &Quark Soliton  &NJL &LC   &\\
\hline
$\delta u$      &0.84     &1.33     &1.09  & 1.17 &1.07    &0.82 &1.17  & \\%
$\delta d$      &-0.23    &0.04     &-0.27 &-0.26  &-0.38   &-0.07 & -0.29& \\%
\hline
$g^t_{b_1 NN}$   &10.31$\sim$17.77&12.43$\sim$21.43&13.1$\sim$22.59&13.78$\sim$23.75&13.97$\sim$24.08&8.58$\sim$14.78&14.07$\sim$24.25&\\%
\hline
$g^t_{h_1 NN}$   &5.28$\sim$9.1 &11.85$\sim$20.43 & 7.09$\sim$12.23&7.87$\sim$13.57&5.96$\sim$10.29&6.49$\sim$11.18&7.61$\sim$13.12 &\\%
\end{tabular}
\end{ruledtabular}
\end{table*}
%


\subsection{Tensor meson couplings}

Since the radiative decay $a_2\to \gamma\pi$ is empirically known
with its width $\Gamma_{a_2\to \pi\gamma}=(0.287\pm 0.03)$ MeV
reported in the Particle Data Group, we estimate the coupling
constant for the interaction $\gamma\pi T$ by using the effective
Lagrangian \cite{giac}
\begin{eqnarray}\label{tensor-r}
&&{\cal L}_{\gamma \pi T}=-i\frac{g_{\gamma\pi
T}}{m^2_0}\tilde{F}_{\alpha\beta}(\partial^\alpha
T^{\beta\rho}-\partial^\beta
T^{\alpha\rho})
\partial_\rho\pi+{\rm H.c}.\ ,\ \ \ \ \
\end{eqnarray}
where
$\tilde{F}_{\alpha\beta}=\frac{1}{2}\varepsilon_{\alpha\beta\mu\nu}F^{\mu\nu}$
is the pseudotensor photon field and the $T^{\alpha\rho}$ is the
tensor meson field, $a_2$.  From the Lagrangian in Eq.
(\ref{tensor-r}), the decay width is given by \cite{giac}
\begin{eqnarray}\label{ten-decay}
\Gamma_{a_2\to \pi\gamma}=\frac{1}{10\pi}\, \left(\frac{g_{\gamma
\pi a_2}}{m^2_0}\right)^2
\left(\frac{m_{a_2}^2-m^2_\pi}{2m_{a_2}}\right)^5\,,
\end{eqnarray}
and we obtain $g_{\gamma \pi a_2}=0.276$.

It is of importance to determine the $TNN$ coupling constant on
the firm ground. This is, however, not an easy task, not only
because information necessary for this is scarce but also because
existing estimates for this are rather diverse in both sides of
theoretical and experiment studies. Therefore, as a guidance to
figure out the $TNN$ coupling constant, we first attempt to use
the TMD hypothesis in the hadron energy-momentum tensor form
factors.

With an assumption of the $f_2(1270)$-pole dominance in the
nucleon and pion energy-momentum tensor form factors,
$<N|\Theta^{\mu\nu}|N>$ and $<\pi|\Theta^{\mu\nu}|\pi>$,
respectively \cite{ysoh,raman}, the effective Lagrangian for the
$TNN$ coupling
\begin{widetext}
\begin{eqnarray}\label{ten00}
{\cal
L}_{TNN}=-2i\frac{g^{(1)}_{TNN}}{M}\bar{N}(\gamma_\lambda\partial_\sigma
+\gamma_\sigma\partial_\lambda)N\,T^{\lambda\sigma}
+4\frac{g^{(2)}_{TNN}}{M^2}\partial_\lambda\bar{N}\,\partial_\sigma
N\,T^{\lambda\sigma},
\end{eqnarray}
\end{widetext}
leads to the following identity,
\begin{eqnarray}\label{tensor-01}
\frac{2}{M}(g^{(1)}_{f_{2NN}}+g^{(2)}_{f_{2NN}})=\frac{g_{f_2\pi\pi
}}{m_f}\,.
\end{eqnarray}
Details in the derivation of Eq. (\ref{tensor-01}) and the
relevant ${\cal L}_{T\pi\pi}$ are given in Appendix C. Thus, the
TMD enables us to determine $g^{(1)}_{f_2NN}=\pm 2.18$ with
$g_{f_2\pi\pi}=\pm 5.9$ estimated from the decay width
$\Gamma_{f_2\to\pi\pi}=156.9^{+3.8}_{-1.2}$ MeV, as
$g^{(2)}_{f_2NN}\approx 0$ chosen by the normalization conditions
of the one-nucleon state with mass, and spin, respectively.
\cite{ysoh,klei,renn}. In a similar way, we obtain
$g^{(1)}_{a_2NN}=\pm 2.43$ and $g^{(2)}_{a_2NN}=0$ with
$g_{a_2KK}=\pm 6.83$ taken from the observed decay width
$\Gamma_{a_2\to K\bar{K}}= (5.3\pm 0.25)$ MeV \cite{giac}. But we
also note that these predictions by the TMD are widely different
from those found in phenomenological studies of meson-, and
photon-induced reactions. Engles derived estimates of
$g^{(1)}_{f_2NN}=6.45$ and $g^{(2)}_{f_2NN}\approx 0$ from the
backward $\pi N$ dispersion relations \cite{engels}. Kleinert and
Weisz determined $g^{(1)}_{a_2NN}=1.54$ and
$g^{(1)}_{a_2NN}\approx$ $-g^{(2)}_{a_2NN}$ from the dispersion
relations for pion photoproduction at threshold \cite{klei}. (Our
coupling constants are related to Refs. \cite{engels,klei} by
$g^{(1,2)}_{f_2NN}$=$G^{(1,2)}_{f_2NN}/4$ in Eq. (\ref{ten00}))
These values for the $f_2NN$ and $a_2NN$ couplings were found to
be consistent with those from the detailed analysis of the Compton
scattering $\gamma N\to \gamma N$ \cite{klei1}. Unlike the VMD and
AVMD, therefore, it is not clear whether the application of the
TMD is effective for the determination of the coupling constant
$a_2NN$, and $f_2NN$ as well. This point was discussed in Ref.
\cite{raman}, where various versions of the TMD were examined to
conclude that such an assumption of the pure $f_2$-pole dominance
as in Eq. (\ref{tensor-01}) may not be correct and that the proper
use of the TMD would require further contribution including an
additional isoscalar piece in the trace $\Theta^\mu_\mu$, which is
much more involved in the violation of the scale invariance, i.e.,
probably dilaton \cite{carr}, or Pomeron \cite{raman} as a
possible candidate. Hence, we regard that, though a viable
hypothesis analogous to the VMD, the validity of the TMD within
the simple pole description is questionable and needs further test
\cite{suzuki}.

We now proceed to determine the tensor meson coupling constants
from those found in phenomenological analyses,
while keeping the SU(3) symmetry as a good approximation to the
tensor meson nonet coupling to the baryon octet \cite{worden}. For
a reliable choice, we refer to consistency check of the SU(3)
relation between the $a_2NN$ and the $f_{2NN}$ couplings,
\begin{eqnarray}\label{su3}
g_{f_{2NN}}^{(1)}=\frac{1}{\sqrt{3}}(4\alpha-1)\,g_{a_2 NN}^{(1)},
\end{eqnarray}
while we let $g^{(2)}_{f_2NN}\approx 0$, and
$g^{(2)}_{a_2NN}\approx 0$, as before. The couplings of the tensor
meson nonet trajectories to the baryons had been investigated to
test the SU(3) symmetry for the residues of the tensor meson
nonet, and there, the ratio $F/D=-1.8\pm 0.2$ was estimated to
agree with that obtained from high energy experiments
\cite{sarma,gross}. Adopting the ratio in Eq. (\ref{su3}) with the
$g_{f_2NN}^{(1)}$ taken from typical models discussed above, we
present  the SU(3) predictions for the coupling constant of the
$a_2NN$ in Table \ref{tb5}.

\begin{table}{}
\caption{SU(3) predictions for the coupling constant of the tensor
meson $a_2$ from existing estimates of the $f_2$ coupling
constant. The value of $g_{f_2NN}^{(1)}$ is taken from other
references as per alphabetical superscript $^a$Ref. \cite{raman},
$^b$Ref. \cite{gold}, $^c$Refs. \cite{ysoh,borie}, and $^d$Ref.
\cite{engels} in order.}
\begin{ruledtabular}
\label{tb5}
\begin{tabular}{ccccccl}
                              & TMD   &  A    & B   &  C &$(\frac{F}{D})_{\rm exp}=-1.8\pm 0.2$    &\\
\hline
$g_{f_2NN}^{(1)}$             &2.18$^a$  &  3.38$^b$ &5.26$^c$   &6.45$^d$    &       &\\%
\hline
$g_{a_2NN}^{(1)}$             &0.39  &  0.6     &0.94  &1.15     & $\alpha=2.67$, $\frac{F}{D}=-1.6$& \\%
$g_{a_2NN}^{(1)}$             &0.47  &  0.73 & 1.14  & 1.4      & $\alpha=2.25$, $\frac{F}{D}=-1.8$&      \\%
$g_{a_2NN}^{(1)}$             &0.54  &  0.84 & 1.3  & 1.6    & $\alpha=2.0$, $\frac{F}{D}=-2.0$&      \\%
\end{tabular}
\end{ruledtabular}
\end{table}

\begin{table*}
\caption{Coupling constants of exchanged mesons. Radiative decay
widths are given in units of keV. The values in the models,
LMR(including the multiplicative factor $\lambda=2.18$), GLV, KM
\cite{mosel}, and NSC \cite{nsc} are taken from the Regge
approach. The values for NN/YN \cite{nn,yn,nn1}, ESC04
\cite{esc04}, and ESC04$_{97}$ \cite{esc04a,esc04f} are extracted
from the one boson exchange potentials.  The decay width in the
parenthesis is the prediction by the chiral unitary model
\cite{roca}.
}
\begin{ruledtabular}\label{tb4}
\begin{tabular}{cccccccccccl}
                       &   LMR   & GLV  &KM  &  NSC  &NN/YN   & ESC04  & ESC04$_{97}$ &Present work &$\Gamma$(keV)&\\
\hline\hline
$g_{\pi NN}/\sqrt{4\pi}$&3.82    & 3.81  &3.78 &3.7    &3.66 & 3.58    &               &3.78    &  &\\%
\hline
$g_{\gamma\pi^\pm\rho}$&         & 0.224 &0.22 &        &      &        &              &0.224 &68$\pm$7&\\%
$g_{\gamma\pi^0\rho}$  &         & 0.224&      &        &      &        &              &0.255 &90$\pm$12&\\%
$g^v_{\rho NN}$        &  2.8    &  3.4 &3.4   &3.16  &2.11  & 2.77    &2.967          &2.6    &    &\\%
$g^t_{\rho NN}$        &$\kappa_\rho$=14.6&$\kappa_\rho$=6.1&6.1&13.338 &17.08&12.31&12.52 &$\kappa_\rho$=6.2&&\\%
\hline
$g_{\gamma\pi^0\omega}$&         & 0.687 &     &        &      &        &                &0.723 &757$\pm$27&\\%
$g^v_{\omega NN}$      &         &15     &     &10.446 &11.96    & 10.683      & 10.36   & 15.6&& \\%
$g^t_{\omega NN}$      &         &0      &     & 3.224  &8.3     &   1.583      &4.2     & 0  &&\\%
\hline
$g_{\gamma\pi^\pm b_1}$&        & 0.187  &0.195&        &      &        &                &0.196 &230$\pm$60&\\%
$g_{\gamma\pi^0b_1}$   &        & $\sqrt{2}g_{\gamma\pi^\pm b_1}$&&   &   &  &           &0.189 & VMD  &\\%
$g^v_{b_1 NN}$        &        &   16.44 &      &     &   &  10.96     &                 & - &   &\\%
$g^t_{b_1 NN}$         &        &   0     &$\geq$7.1&     &    &           &              &-14&&\\%
\hline
$g_{\gamma\pi^\pm a_1}$ &        &        &0.33 &        &      &        &               &0.316 &640$\pm$246&\\%
$g^v_{a_1 NN}$          &        &        &7.1  &     &   & 9.014          &             & 6.7 &  &\\%
\hline
$g_{\gamma\pi^0h_1}$   &         &       &      &        &      &        &                &0.405 & (837$\pm$134) &\\%
$g^t_{h_1 NN}$         &        &        &      &     &    &           &                  &-9&&\\%
\hline
$g_{\gamma\pi^\pm a_2}$ &        &       &      &        &      &        &                &0.276 &287$\pm$30&\\%
$g^{(1)}_{a_2 NN}$      &        &       &      &  1.573 &      &     &                & 1.4 &&\\%
$g^{(2)}_{a_2 NN}$      &        &       &      &        &      &     &             & 0  &&\\%
%
\end{tabular}
\end{ruledtabular}
\end{table*}
%


Accordingly, we find that the values $g^{(1)}_{f_2NN}=6.45$ and
$g^{(1)}_{a_2NN}=1.54$ determined in Refs. \cite{klei,klei1} are
in good agreement with the SU(3) predictions
with the case of $\alpha=2.25$ or $\alpha=2.0$ in column C of
Table \ref{tb5}. Furthermore these values are also supported by
the value $g^{(1)}_{a_2NN}=1.57$ extracted from the Nijmegen
soft-core YN potential \cite{nsc,nagel} and its extended version
\cite{rijken}. In this work we favor to choose the SU(3) values
$g^{(1)}_{a_2NN}=1.4$ and $g^{(2)}_{a_2NN}\approx 0$ from Table
\ref{tb5} as a median value for the present calculation. We
consider that the results should be valid to what extent the
symmetry is a good approximation to the couplings of the tensor
meson nonet to baryons. In the next section we calculate
differential cross section and spin polarizations using
$g_{a_2NN}^{(1)}=1.4$ and $g_{a_2NN}^{(2)}=0$. But we also
consider the case of $g_{a_2NN}^{(2)}\approx-g_{a_2NN}^{(1)}$ by
taking $g_{a_2NN}^{(2)}=-1.2$ \cite{klei1} for an exploratory
study and show that numerical evidences support the choice
$g_{a_2NN}^{(2)}=0$ rather than
$g_{a_2NN}^{(2)}\approx-g_{a_2NN}^{(1)}$ in the analysis of these
observables.

\section{Results and discussion}

In this section we present the numerical results of the four
photoproduction processes, $\gamma p\to \pi^+ n$, $\gamma n\to
\pi^- p$, $\gamma p\to \pi^0 p$ and $\gamma n\to \pi^0 n$, with
the coupling constants chosen in Table \ref{tb4}.

\subsection{Charged pion photoproduction}

\subsubsection{Differential cross section}

The dynamical feature of the charged pion process is characterized
by the sharp rise at the very forward angle $-t\leq m_\pi^2$ in
the differential cross section as well as the spin polarization
asymmetry.

\begin{figure}[htbp]
\includegraphics*[width=8cm]{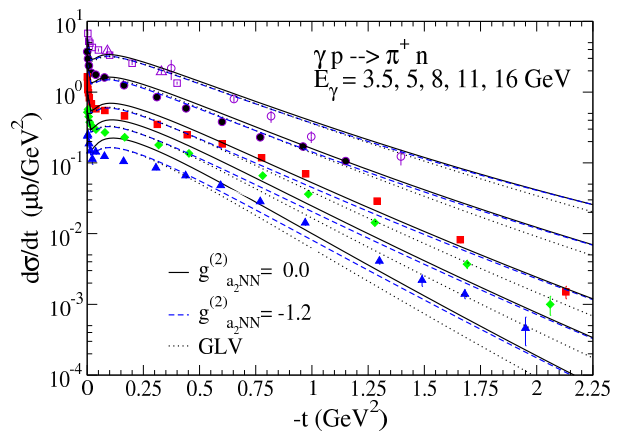}
\caption[]{(Color online)Differential cross sections for $\gamma
p\to \pi^+ n$ at photon energies $E_\gamma=3.5, 5,\,8,\,11$, and
$16$ GeV, respectively. Solid curves result from the gauge
invariant $(\pi+b_1)+a_1+(\rho+a_2)$ exchanges with
$g^{(1)}_{a_2NN}=+1.4$ and $g^{(2)}_{a_2NN}=0$, while dashed
ones(blue) from $g^{(1)}_{a_2NN}=+1.4$ and $g^{(2)}_{a_2NN}=-1.2$.
Dotted curves result from the GLV model with the gauge invariant
$\pi+\rho$ exchanges. The data are taken from Refs. \cite{heide68}
(open squares), \cite{joseph67} (open triangles), \cite{baryam67}
(open circles) and \cite{boyarski68} (filled circles, filled
squares, filled diamonds and filled triangles). }
\label{fig:1dsdtallnew}
\end{figure}

\begin{figure}[htbp]
\includegraphics*[width=4cm]{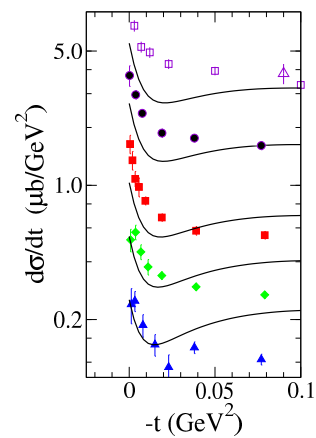}
\caption[]{(Color online) Forward peaks in the region $|t|\approx
m_\pi^2$ magnified from Fig. \ref{fig:1dsdtallnew} with the same
notation for the solid curve. } \label{fig:1dsdtalllowtc}
\end{figure}

\begin{figure}[htbp]
\includegraphics*[width=8cm]{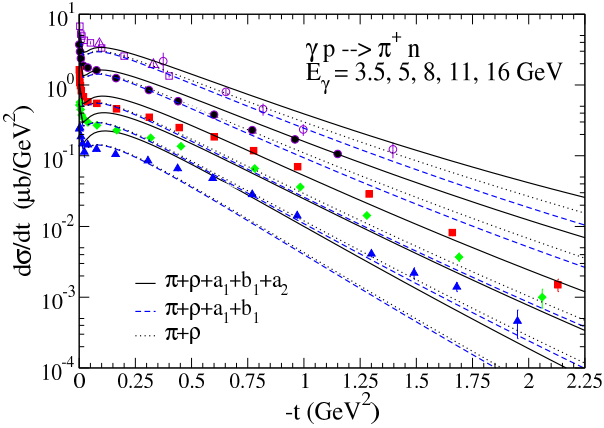}
\caption[]{(Color online) Role of the tensor meson $a_2$ Regge
pole in the differential cross section for $\gamma p\to \pi^+ n$.
The solid line is our model prediction from the
$(\pi+b_1)+a_1+(\rho+a_2)$ exchanges with
$g^{(1)}_{a_2NN}(g^{(2)}_{a_2NN})=1.4(0)$, and the blue dashed
line is from $\pi+b_1+a_1+\rho$. The dotted line is from the
primary $\pi+\rho$ exchanges.} \label{fig:1dsdtalla2role}
\end{figure}

Figure \ref{fig:1dsdtallnew} shows the forward angle differential
cross sections for the $\gamma p\to \pi^+n$ process at five photon
energies $E_\gamma=$3.5, 5, 8, 11 and 16 GeV. The solid curve
results from the present work. The cross section is not sensitive
to the choice between $g^{(2)}_{a_2NN}=0$ and
$g^{(2)}_{a_2NN}=-1.2$ as shown. As the unnatural parity exchange,
the contribution of the axial meson is found insignificant in the
differential cross section. Compared to the GLV model (the dotted
curve), it is clear that the exchange of the $a_2$ in the present
model makes improved the $t$-dependence of the cross section as
the energy and the momentum transfer $-t$ increase. The forward
peaks near $|t|\approx m_\pi^2$ are reproduced in Fig.
\ref{fig:1dsdtalllowtc} to show the role of the nucleon Born term
in this region. In order for the clear differentiation between the
model prediction with and without the $a_2$ exchange, we present
Fig. \ref{fig:1dsdtalla2role} for future experimental test.

\begin{figure}[htbp]
\includegraphics*[width=8cm]{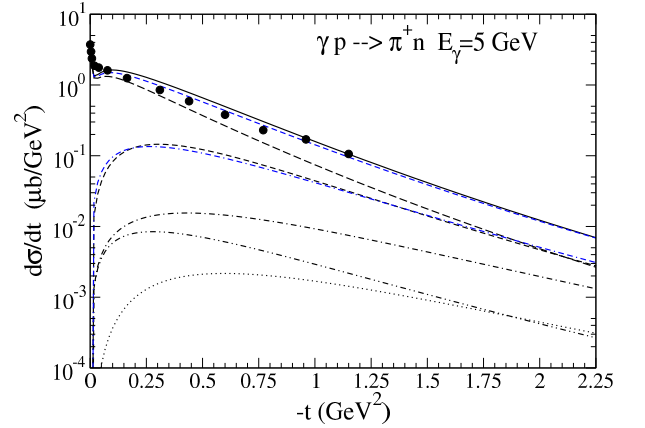}
\caption[]{(Color online) Contribution of each meson exchange to
the differential cross section for $\gamma p\to \pi^+ n$ at
$E_\gamma=5$ GeV. The solid line is our model prediction from the
$(\pi+b_1)+a_1+(\rho+a_2)$ exchanges with
$g^{(1)}_{a_2NN}(g^{(2)}_{a_2NN})=1.4(0)$, and the corresponding
one with the $g^{(1)}_{a_2NN}(g^{(2)}_{a_2NN})=1.4(-1.2)$ is given
by the dashed line(blue). The long dashed line(black) is from
$\pi$ exchange. The short dashed line(black) is from the $\rho$
exchange. The dash-dash-dotted line(blue) is from the $a_2$
exchange with $g^{(1)}_{a_2NN}(g^{(2)}_{a_2NN})=1.4(0)$, and the
dash-dotted line is from  the $a_2$ exchange with
$g^{(1)}_{a_2NN}(g^{(2)}_{a_2NN})=1.4(-1.2)$. The dot-dot-dashed
line is from $a_1$ exchange, and the dotted line from $b_1$
exchange.} \label{fig:1dsdt5a2role}
\end{figure}

In Fig. \ref{fig:1dsdt5a2role} the contribution of each meson
exchange to the cross section is displayed for the $\gamma
p\to\pi^+n$ process at $E_\gamma=5$ GeV. The pion exchange gives
the leading contribution as depicted by the long dashed line. The
$a_2$ exchange with $g_{a_2NN}^{(1)}=1.4$ and $g_{a_2NN}^{(2)}=0$
gives the contribution comparable to that of the $\rho$ exchange,
both of which are smaller than the pion exchange by an order of
magnitude. However, the case of $g_{a_2NN}^{(2)}=-1.2$ falls off
the $a_2$ contribution to be nearly the same contribution of
$a_1$, which is smaller than the pion exchange by two orders of
magnitude.

\begin{figure}[htbp]
\includegraphics*[width=8cm]{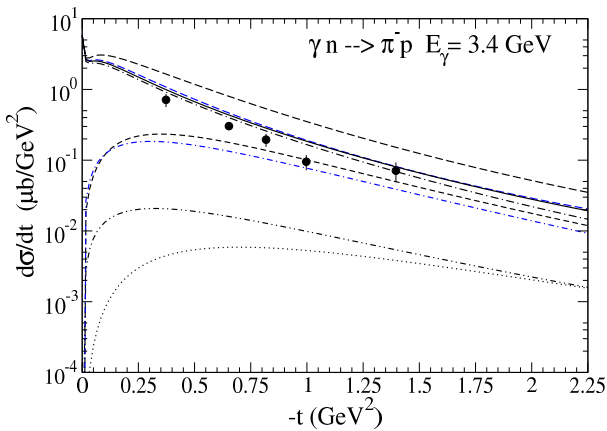}
\caption[]{(color online) Differential cross section for $\gamma
n\to \pi^-p$ at $E_\gamma=3.4$ GeV. Notations are the same as in
Fig.  \ref{fig:1dsdt5a2role} except for the dot-dashed line from
the GLV model calculation. For the given $g_{a_2NN}^{(1)}=1.4$,
the solid line results from the $(-\pi+b_1)-a_1+(\rho-a_2)$
exchanges with $g_{a_2NN}^{(2)}=0$, while the dashed one(blue) is
from $g_{a_2NN}^{(2)}=-1.2$. The data are taken from Ref.
\cite{baryam67}. } \label{fig:5dsdtpimp}
\end{figure}

We present the differential cross section for the $\gamma n\to
\pi^-p\ $ process at $E_\gamma=3.4$ GeV in Fig.
\ref{fig:5dsdtpimp}. The constant phase for all exchanged mesons
required by the EXD is consistent with data, as discussed in Eq.
(\ref{born-}). Each meson exchange gives the same contribution as
the case of the $\gamma p\to\pi^+n$ process, but with the opposite
signs in the $\pi$, $a_1$ and $a_2$ exchanges by the $G$ parity
argument.

\begin{figure}[htbp]
\includegraphics*[width=8cm]{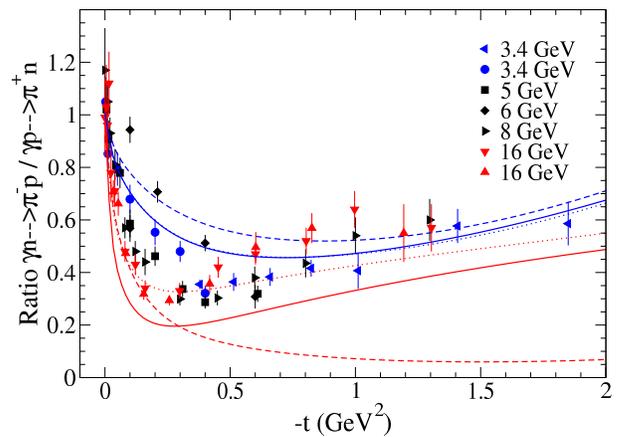}
\caption[]{(Color online) Ratio $R=\frac{d\sigma/dt(\gamma n\to
\pi^-p)}{d\sigma/dt(\gamma p\to \pi^+n)} $ at $E_\gamma=3.4,\,
5,\,6,\,8,\,$ and $16$ GeV. Model predictions are given for the
two cases $E_\gamma=3.4$ (upper three blue lines) and 16 GeV
(lower three red lines). Solid lines are from the present work
with $g_{a_2NN}^{(1)}=1.4$ and $g_{a_2NN}^{(2)}=0$, while dashed
ones are from $g_{a_2NN}^{(1)}=1.4$ and $g_{a_2NN}^{(2)}=-1.2$.
Dotted lines are from the GLV model. The result at $E_\gamma$=16
GeV supports $g_{a_2NN}^{(2)}=0$. The data are taken from Refs.
\cite{heide68} (filled squares, filled circles and filled
diamonds), \cite{baryam67} (filled left-triangles),
\cite{boyarski68} (filled down-triangles, filled right-triangles),
and \cite{sherden73} (filled up-triangles). }
\label{fig:rdsdtourglv}
\end{figure}

The measurement of the ratio
\begin{eqnarray}
R=\frac{d\sigma/dt(\gamma n\to \pi^-p)}{d\sigma/dt(\gamma p\to
\pi^+n)}
\end{eqnarray}
is of significance to test the validity of the Regge formalism
together with the coupling constant and phase chosen for each
meson exchange. In Fig. \ref{fig:rdsdtourglv}  the dominance of
the nucleon Born term in both processes leads the ratio to be
unity at $-t\approx 0$. The marked minimum just below $-t\approx
0.5$ GeV$^2$ may reach by the opposite interference patterns
between the $\pm\pi+\rho\pm a_2$ contributions to each other
process. It should be noted that the dashed line for the ratio at
$E_\gamma=$16 GeV with
$g_{a_2NN}^{(1)}(g_{a_2NN}^{(2)})=1.4(-1.2)$ is wide out of the
data points and, hence, fails to reproduce the measurement. Such
numerical evidence supports $g_{a_2NN}^{(2)}=0$ as compared to the
solid line (red) at the same energy.

\subsubsection{Spin polarization asymmetries}

Single spin polarizations are analyzed for the polarized photon
and the target nucleon.

The photon polarization symmetry $\Sigma$ is defined as
\begin{eqnarray}
\Sigma&=&
\frac{d\sigma_{\perp}-d\sigma_{\parallel}}{d\sigma_{\perp}
+d\sigma_{\parallel}}\ ,
\end{eqnarray}
where $d\sigma_{\perp}(d\sigma_{\parallel})$ is the differential
cross section for the photon polarization along $x(y)$ axis with
the $z$-axis denoted by the direction of the photon propagation in
the reaction plane. In terms of the helicity amplitudes
\cite{levy} the $\Sigma$ measures the asymmetry between the
natural and the unnatural parity exchange so that the
contributions of the axial mesons $a_1$ and $b_1$, though
negligible for the differential cross section, are to be exposed.

\begin{figure}[htbp]
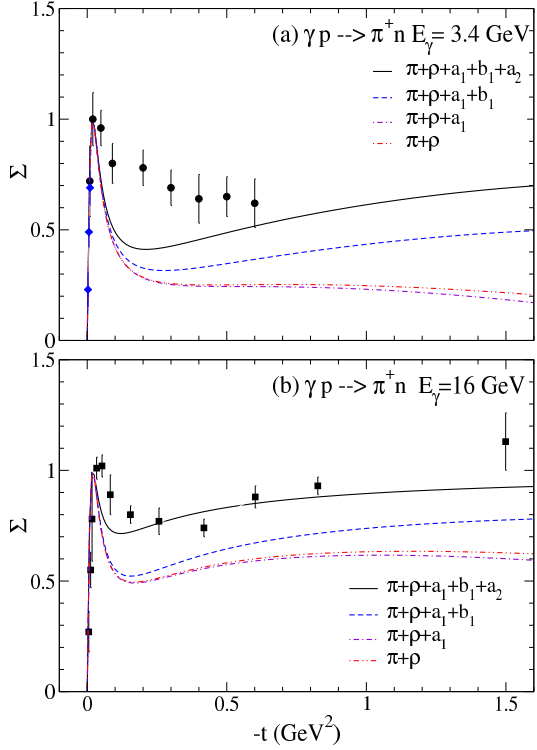

\includegraphics*[width=7.0cm]{1phoasy3p4tnew.eps}
\includegraphics*[width=7.0cm]{1phoasy16tnew.eps}
\caption[]{(Color online) Photon polarization asymmetry for
$\gamma p\to \pi^+n$ at (a) $E_\gamma=3.4$ GeV,
and (b) $E_\gamma=16$ GeV. The solid line results from
$(\pi+b_1)+a_1+(\rho+a_2)$ exchanges with $g_{a_2NN}^{(2)}=0$. The
dashed line is from $(\pi+b_1)+a_1+\rho$ exchanges. The
dash-dotted line is from $\pi+a_1+\rho$ exchanges. The
dash-dot-dotted line is from $\pi+\rho$ exchanges. The data are
taken from Refs. \cite{sherden73} (filled squares),
\cite{burfein70} (filled diamonds), and \cite{gewen69} (filled
circles and filled triangles). } \label{fig:1phoasy}
\end{figure}

\begin{figure}[htbp]
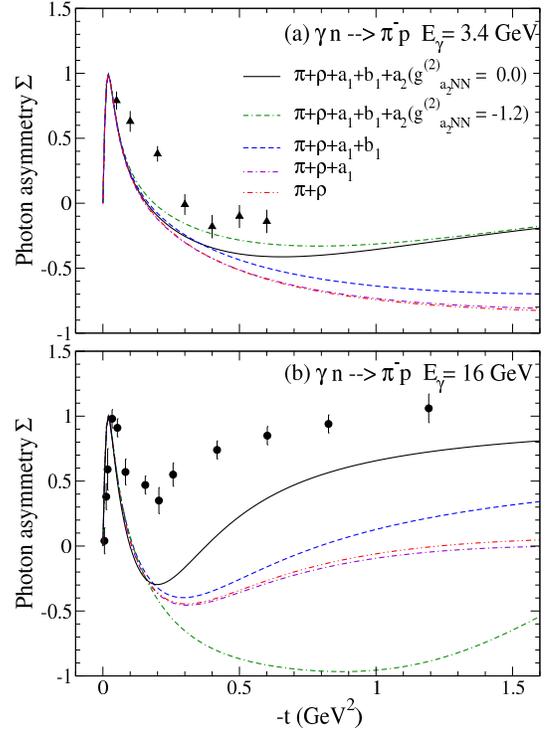

\includegraphics*[width=7.0cm]{5phoasy3p4rtnew.eps}
\vfill
\includegraphics*[width=7.0cm]{5phoasy16rtnew.eps}
\caption[]{(Color online) Photon polarization asymmetry for
$\gamma n\to \pi^-p$ at (a) $E_\gamma=3.4$ GeV (b) $E_\gamma=16$
GeV. Notations are the same as in Fig.  \ref{fig:1phoasy}. The
solid line is from the present model with
$g_{a_2NN}^{(2)}$= 0. The dash-dash-dotted line(green) results
from $g_{a_2NN}^{(2)}= -1.2$, showing the failure to reproduce
data points at $E_\gamma=16$ GeV. The data are taken from Refs.
\cite{burfein73}.}\label{fig:5phoasyr}
\end{figure}

Figure \ref{fig:1phoasy} shows the photon polarization $\Sigma$ in
the $\gamma p\to\pi^+n$ process with $g_{a_2NN}^{(2)}=0$ in accord
with the numerical consequence in Fig. \ref{fig:rdsdtourglv}. The
case of $\gamma n\to\pi^-p$ is presented in Fig.
\ref{fig:5phoasyr} which is further added in proof for
$g_{a_2NN}^{(2)}=0$, as shown by the numerical consequence at
$E_\gamma=16$ GeV.

In both charged pion processes, the rapid rise of the $\Sigma$
near threshold is reproduced by the nucleon Born term. Then the
$\Sigma$ is expected to approach to unity by the dominance of the
natural parity exchange as the energy increases. It should be
stressed that the role of the $a_2$ exchange is crucial for the
better description of the $\Sigma$. Without the $a_2$ exchange our
results in the $\Sigma$ (the dashed lines) are nearly the same as
those of the GLV model. We also indicate that our model exhibits
the sizable contribution of the axial meson $b_1$ in these
measurements.

\begin{figure}[htbp]
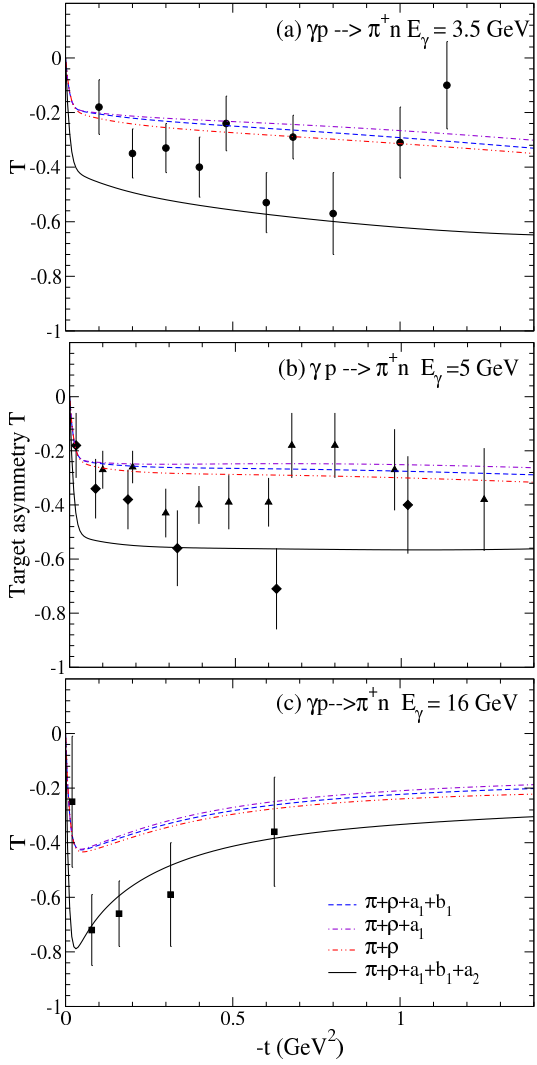

\includegraphics*[width=7cm]{1tarasy3p5tnew.eps}
\includegraphics*[width=7cm]{1tarasy5tnew.eps}
\includegraphics*[width=7cm]{1tarasy16tnew.eps}
\caption[]{(Color online) Target polarization asymmetry for
$\gamma p\to \pi^+n$ at (a) $E_\gamma=3.4$ GeV, (b) $E_\gamma=5$
GeV and (c) $E_\gamma=16$ GeV, respectively in the present work.
The solid line results from $(\pi+b_1)+a_1+(\rho+a_2)$ exchanges
with $g_{a_2NN}^{(2)}=0$. The data are taken from Ref.
\cite{genzel75} (filled circles and filled triangles) and
\cite{more70} (filled diamonds and filled squares) }
\label{fig:1tarasy}
\end{figure}

The target polarization asymmetry ($T$) is defined by
\begin{eqnarray}
T=\frac{d\sigma_{\uparrow}-d\sigma_{\downarrow}}{d\sigma_{\uparrow}
+d\sigma_{\downarrow}}\
, \label{eqpol}
\end{eqnarray}
which measures the asymmetry of the spin polarization of the
target nucleon parallel and antiparallel to the direction
$\frac{\vec{k}\times\vec{q}}{|\vec{k}\times\vec{q}|}$ in the
center-of-mass system \cite{bgyu-jpg}. With the sharp rise at
threshold due to the nucleon Born term, our results exhibit the
decisive role of the $a_2$ exchange in the target polarization, in
particular at $E_\gamma$=16 GeV, as shown in Fig.
\ref{fig:1tarasy}. We note that the results without the $a_2$
exchange are nearly the same as those of the GLV model calculation
(similar to those lines except for the solid one) \cite{guid}.

\subsection{Neutral pion photoproduction}

In the absence of the pion exchange from the neutral case, the
dynamics of the production mechanism is solely determined by the
exchange of $\omega+\rho^0+b_1+h_1$ Regge poles, unless the cuts
are considered.

\subsubsection{Differential cross section and spin polarization
asymmetry}

The appearance of the deep dip characterizes the feature of the
$\gamma p\to \pi^0 p$ process around $-t\approx$ 0.5 GeV$^2$ in
the differential cross section as well as in the photon
polarization. We follow the procedure of Ref. \cite{guid} for the
dip-generating mechanism by the nonsense wrong signature zero of
the dominating Regge trajectory.

In comparison with the coupling strengths between exchanged
mesons, the $\omega$ exchange would predominate in the process
with the large coupling constants for electromagnetic and strong
interactions. We, therefore, describe the dip region of the cross
section with the nonsense zero of the $\omega$ trajectory, i.e.,
$\alpha_\omega(t) \approx 0$, which occurs precisely at the place
the dip is observed in the experiment. We take the phase of the
$\omega$ exchange to be non-degenerate,
$\frac{1}{2}(-1+e^{-i\pi\alpha_\omega})$, while that of the $\rho$
should be kept degenerate, $e^{-i\pi\alpha_\rho}$, to fill up the
singularity of the amplitude at $-t\approx 0.5$ GeV$^2$ caused by
the nonsense zero value of the $\alpha_\omega$. As an additional
dip-filling mechanism, the axial meson exchanges $b_1$ and $h_1$
also contribute with the constant phase in common.

\begin{figure}[htbp]
\includegraphics*[width=8cm]{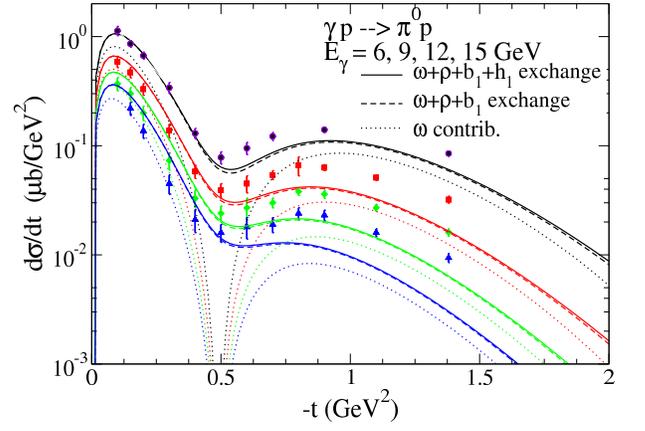}
\caption[]{ (Color online) Cross sections for $\gamma p\to \pi^0
p$ at $E_\gamma=6,\,9,\,12$, and $15$ GeV from the
$\omega+\rho+b_1+h_1$ exchanges in the present model. The data are
taken from Ref. \cite{anderson71}.}
 \label{fig:8dsdtall}
\end{figure}

The cross sections for the $\gamma p\to\pi^0 p$ process at
$E_\gamma=6,9,12$ and 15 GeV are shown in Fig.
\ref{fig:8dsdtall}. We find, however, that the
$\omega+\rho+b_1+h_1$ exchanges with the coupling constants in
Table \ref{tb4} are in less agreement with the cross section data
above the region $-t\approx$ 0.5 GeV$^2$ as the photon energy
increases.
%
%
Within the context of the present approach where the gross feature
of the cross section depends mainly on the singularity of the
$\omega$ exchange with the leading trajectory $\alpha_\omega$, our
model shows deficiency in reproducing the cross section, in
particular, over the region $-t\approx 0.5$ GeV$^2$ at the higher
photon energies, because of the more rapid falloff in the axial
meson contributions due to their trajectories lying lower than
that of the $\omega$ \cite{donn}.

\begin{figure}[t]
\includegraphics*[width=8cm]{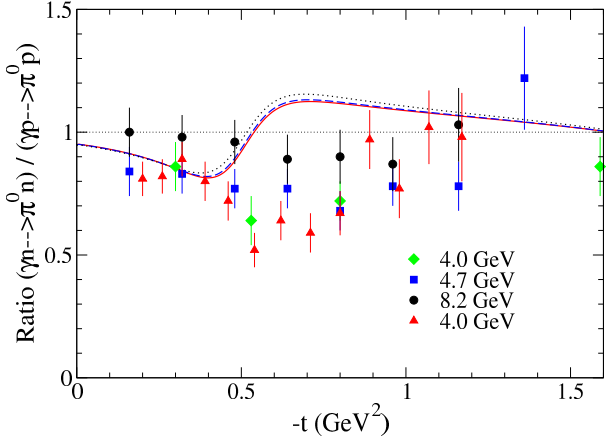}
\caption[]{(Color online) Ratio $R=\frac{d\sigma/dt(\gamma n\to
\pi^0n)}{d\sigma/dt(\gamma p\to\pi^0p)}$ for the differential
cross sections.  The data are taken from Refs. \cite{bolon71}
(filled diamonds), \cite{osborne72} (filled squares, filled
circles) and \cite{braun73} (filled triangles). }
\label{fig:11rdsdtpi0}
\end{figure}

The ratio for the neutral pion process is presented in Fig.
\ref{fig:11rdsdtpi0},
\begin{eqnarray}
 R=\frac{d\sigma/dt(\gamma n\to \pi^0n)}{d\sigma/dt(\gamma p\to
\pi^0p)}\,,
\end{eqnarray}
which shows a slight dip structure in the experimental data.
Figure \ref{fig:11rdsdtpi0} has obtained with all the phases of
the Regge poles, $\rho$, $b_1$, and $h_1$ chosen as constant for
the $\gamma n\to\pi^0n$ process, while for the $\gamma p\to\pi^0p$
process the respective phases of these poles are taken as
discussed above. The solid line draws the ratio at $E_\gamma=4$
GeV. The dashed one at $E_\gamma=4.7$ GeV and dotted one at
$E_\gamma=8.2$ GeV, respectively. The results are not good enough
to explain the data, in particular, over the region $-t\approx$
0.5 GeV$^2$. This may reflect the deficiency of the model
prediction for the cross section in the region as discussed in
Fig. \ref{fig:8dsdtall}. From the isospin symmetry and the
negligible role of the axial meson in the cross section, we write
the ratio as $R\sim$ $
\frac{|\omega-\rho^0|^2}{|\omega+\rho^0|^2}\,$ symbolically, and
expect that the ratio eventually reaches unity by the dominance of
the $\omega$, as the data points do at $E_\gamma=8.2$ GeV.

\begin{figure}[htbp]
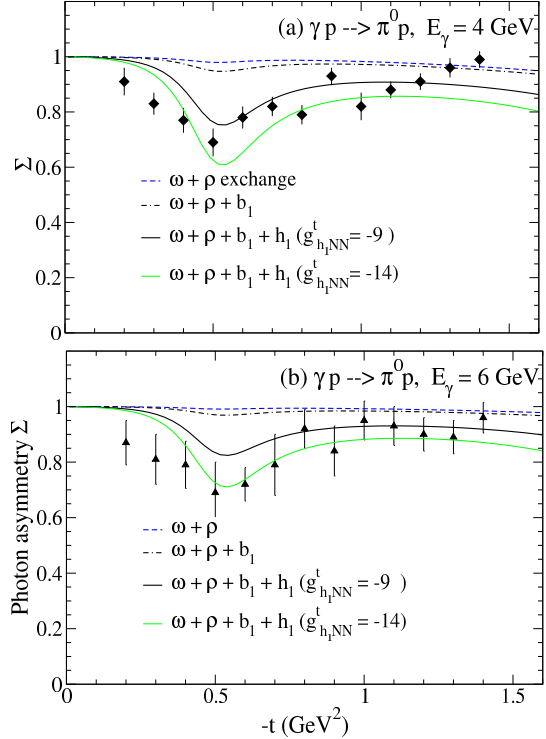

\includegraphics*[width=7cm]{8phoasy4tnn.eps}
\vfill
\includegraphics*[width=7.1cm]{8phoasy6tnn.eps}
\caption[]{(Color online) Photon polarization asymmetry for
$\gamma p\to \pi^0p$ at (a) $E_\gamma=4$ GeV, (b) $E_\gamma=6$
GeV. The solid curve results from $\omega+\rho+b_1 +h_1 $
exchanges in the present model. The lower solid line(green)
results from the change of $g_{h_1NN}^t=-9$ to $-14$. The data are
taken from Ref. \cite{anderson71}.} \label{fig:8phoasy}
\end{figure}

The photon polarization $\Sigma$ is presented in Fig.
\ref{fig:8phoasy}. The cross section, the ratio $R$, and the
photon polarization asymmetry $\Sigma$ for the neutral pion case
are sensitive to a change of the phase for the Regge trajectory.
Since the $\Sigma$ measures the asymmetry between the natural and
the unnatural exchange, i.e., $\Sigma$ $\simeq$
$\frac{|\omega+\rho^0|^2-|b_1+h_1|^2}{|\omega+\rho^0|^2+|b_1+h_1|^2}
$ written as before, we attempt to modulate the depth of the dip
by the contribution of the axial meson negative to the dominating
$\omega$ exchange and find that the isoscalar $h_1$ meson plays
the role as shown in Fig. \ref{fig:8phoasy}. It is interesting to
see that the $h_1$ gives much larger contribution than the $b_1$.
Recall that the solid line (black) is obtained by using the
$g_{h_1NN}^t=-9$ estimated from $<q_\perp^2>\approx $ 1 GeV$^2$ in
Eq. (\ref{tensorcc}) as given in Table \ref{tb4}. But simple guess
from the uncertainty principle $\Delta x\,\Delta q\sim \hbar$
gives much smaller value for the quark momentum squared inside
nucleon. The chiral quark model estimate of $<q_\perp^2>=$ 0.224
GeV$^2$ in Ref. \cite{waka}, for instance,
leads to even the larger values for both axial meson coupling
constants. The lower solid line (green color) in Fig.
\ref{fig:8phoasy} shows the case with $g_{h_1 NN}^t=-14$ chosen as
much as the $b_1$. It is legitimate to point out that the result
from the GLV model is not valid for the present case because the
vector coupling of the axial meson $b_1$ they used is not allowed
by charge conjugation.

\begin{figure}[tbp]
\includegraphics*[width=8cm]{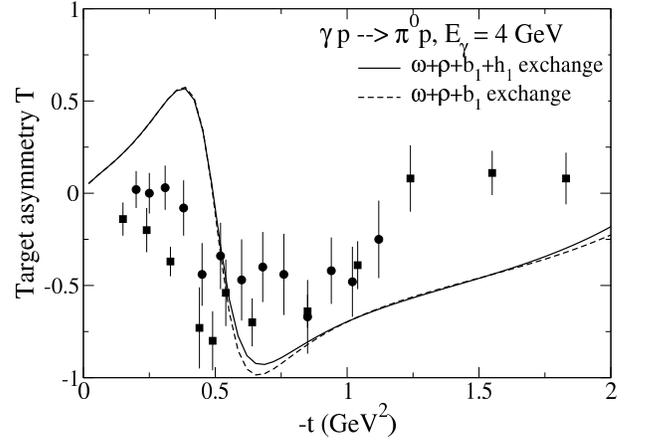}
\caption[]{(Color online) Target polarization asymmetry at
$E_\gamma=$ 4 GeV in the present work. The data are taken from
Refs. \cite{booth72} (filled squares) and \cite{beinlein73}
(filled circles). } \label{fig:8tarasy4}
\end{figure}

\begin{figure}[tbp]
\includegraphics*[width=8cm]{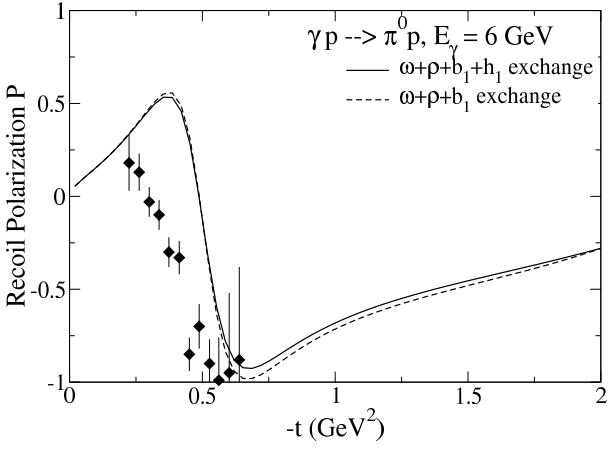}
\caption[]{(Color online) Recoil polarization asymmetry at
$E_\gamma=$ 6 GeV in the present work. The data are taken from
Ref. \cite{deutsch72}. } \label{fig:8recpol6}
\end{figure}

In Figs.  \ref{fig:8tarasy4} and \ref{fig:8recpol6} we show the
target and recoil polarization asymmetries at $E_\gamma$=4 GeV and
at $E_\gamma$=6 GeV, respectively. These observables are by
definition related to each other through the model-independent
inequality \cite{sibi},
\begin{eqnarray}
|P-T|\leq 1-\Sigma\,,
\end{eqnarray}
which predicts that they are approximately equal in case of
$\Sigma\approx$ 1. Indeed, the relation,
\begin{eqnarray}
P-T=4\pi\sqrt{-t}\,{\rm Im}[f_2^{11}f_2^{01\,*}]\,,
\end{eqnarray}
given in terms of the TCHA proves that $P=T$ in the present
framework, because the $f_2^{11}=0$ from the fact that the ${\cal
A}_3=0$ with null contributions of the axial and vector meson
exchanges in general, as given in Appendix  A. But we need more
data in the case of the recoiled polarization to verify the above
relation as well as validity of the given model.

We summarize the motivation of the present work and the result we
have accomplished here as follows; the present work is initiated
by the advantages of the SCHA over the TCHA, which opens the
possibility of incorporating the Regge poles in the Born
approximation amplitude \cite{levy}. This point is worth
rephrasing because we thus have an effective theory which enables
us to continually work with the Born amplitude from threshold to
high energy region in a consistent manner. The theoretical
consideration remaining there is an addition of the chiral loop
contribution near threshold \cite{chpt}, or the replacement of the
fixed-$t$ pole by the Regge pole at high energy \cite{levy,guid}.
In this sense it is of value to establish the coupling strength of
the exchanged particle in an acceptable range of physical values
throughout the energy region and, hence, to have the Regge theory
basically free of parameters, as we have elaborated on here. To
this end we introduce the tensor meson $a_2$, and the axial meson
$h_1$ to the existing Regge model of the $\pi+\rho$ exchanges at
the price of its own simplicity. Though not apparent in the
present process, it is likely that the difference of the
meson-baryon coupling constants between our model and those in
Refs. \cite{levy,guid} becomes significant in the case of kaon
photoproduction. This will be discussed elsewhere.

In the charged pion photoproduction, the inclusion of the $a_2$
exchange in the primary $\pi+\rho+ \rm nucleon\ \rm Born\ \rm
term$ yields an improved $t$-dependence of the cross section up to
$-t\approx $2 GeV$^2$ as well as the photon polarization. We
demonstrate the validity of the coupling constants chosen here by
showing the numerical results in better agreement with data. For
the neutral case, the exchange of the axial meson $h_1$ is newly
found to give a nontrivial contribution to the photon
polarization, while the contribution of the $b_1$ exchange appears
in minor roles.

\begin{acknowledgments}

This work was supported by the Korea Research Foundation Grant
funded by the Korean Government (KRF-2008-313-C00205).

\end{acknowledgments}

\section*{APPENDIX A}

It is worth noting the difference between the TCHA and the SCHA
within the Regge formalism.
For the two body reaction proceeded via the particle exchange in
the $t$-channel, $1+\bar{3}\to \bar{2}+4$, the Regge pole of a
meson exchange in the TCHA is typically given by
\begin{eqnarray}\label{tcha}
f_i^{\mu_{13},\mu_{24}}(s,t)=
\beta(t)\frac{1+\tau
e^{-i\pi\alpha(t)}}{\sin\pi\alpha(t)}\left(\frac{s}{s_0}\right)^{\alpha(t)}\,.
\end{eqnarray}
The $\beta(t)$ is the factorized residue function with the
respective helicity changes denoted by the superscript
$\mu_{ij}=(\mu_i-\mu_j)$. 
In order to be free of kinematical singularities the residue
involves the complicated kinematical $t$ factors,
$\alpha(t),\,(\alpha(t)+1),\,\cdots ,$ to suppress the zeros of
$\sin\pi\alpha(t)$ at negative integer values of $\alpha(t)$
\cite{sibi}. To account for the coupling strengths at the
interaction vertices, the residue usually contains some parameters
to be fitted to empirical data. Since the residue is not in the
form of the vertex coupling usually given by the diagrammatic
technique, those parameters are hard to imply the coupling
constants in the usual sense.

On the other hand, through crossing of helicity amplitudes $1+2\to
3+4$ for the process in the $s$-channel exchange, the Regge pole
in the SCHA corresponding to Eq. (\ref{tcha}) is in general
written as \cite{cohen,bell,coll}
\begin{eqnarray}\label{scha00}
H_i^{\mu_{34},\mu_{12}}(s,t)=\left(\frac{-t}{s_0}\right)^{\frac{1}{2}(n+x)}\gamma(t)
\,\frac{1+\tau
e^{-i\pi\alpha(t)}}{2\sin\pi\alpha(t)}\left(\frac{s}{s_0}\right)^{\alpha(t)},
\nonumber\\
\end{eqnarray}
where the $\gamma(t)$=$\gamma_{13}(t)\gamma_{24}(t)$ is the
factorized residue function with the net helicity flip
$n=|(\mu_1-\mu_2)-(\mu_3-\mu_4)|$. The kinematical $t$ factor in
front of the residue comes from the half-angle factors,
$\left(\frac{s}{s_0}\frac{1-\cos\theta}{2}\right)$ $\to$
$\left(\frac{-t}{s_0}\right)$ and
$\left(\frac{1+\cos\theta}{2}\right)$ $\to$ 1 in the limit $s\gg$
$4M^2$ and small $t$ \cite{coll}. In this case,
$(n+x)=|\mu_1-\mu_3|+|\mu_2-\mu_4|$ and the factor in power of $x$
is called an evading factor usually neglected for the case of
equal masses between the particles. In the SCHA the essential
$t$-singularities are preserved in the half-angle factors, though
care must be taken for the evading factor which is introduced by
the fact that the Regge pole is a definite parity state in the
$t$-channel with its residue factorizing in terms of $t$-channel
helicities \cite{coll}.

For the photoproduction with the notation $1(\gamma)+2(N)\to
3(\pi)+4(N)$ above, the reggeized SCHA in Eq. (\ref{scha-00})
meets with the form in Eq. (\ref{scha00}) where  the
$t$-singularities correspond to the half-angle factors, and the
helicity changes are given by $H_1^{\frac{1}{2},\frac{3}{2}}$,
$H_2^{\frac{1}{2},\frac{1}{2}}$, $H_3^{-\frac{1}{2},\frac{3}{2}}$,
and $H_4^{-\frac{1}{2},\frac{1}{2}}$ or by $H_1^{(-,-)}$,
$H_2^{(-,+)}$, $H_3^{(+,-)}$, and $H_4^{(+,+)}$ in terms of the
final and initial nucleon helicities in order in the superscript
\cite{walker}.

The TCHA, $f_i^{\mu,\mu'}$, is related to the CGLN-invariant
amplitudes \cite{levy,sibi}. Given explicitly,
\begin{eqnarray}
&&f_1^{01}={\cal A}_1-M_+\,{\cal A}_4\ ,\nonumber\\%
&&f_2^{01}={\cal A}_1+(\,t-M_-^2\,)\,{\cal A}_2-M_-\,{\cal A}_3\ ,\nonumber\\%
&&f_1^{11}=M_+\,{\cal A}_1-\,t\,{\cal A}_4\ ,\nonumber\\%
&&f_2^{11}=-M_-\,{\cal A}_1+\,t\,{\cal A}_3\ ,
\end{eqnarray}
and the differential cross section and the photon polarization in
this channel are
\begin{eqnarray}
\frac{d\sigma}{dt}&=&\frac{1}{32\pi}\left[ \frac{
t|f_1^{01}|^2-|f_1^{11}|^2}{t-M_+^2}
+\frac{t|f_2^{01}|^2-|f_2^{11}|^2}{t-M_-^2}\right],\ \ \label{obs1}
\\%
\frac{d\sigma}{dt}\Sigma&=&\frac{1}{16\pi}\left[ \frac{
t|f_1^{01}|^2-|f_1^{11}|^2}{t-M_+^2}
-\frac{t|f_2^{01}|^2-|f_2^{11}|^2}{t-M_-^2}\right],\ \
\end{eqnarray}
with $M_{\pm}=(M'\pm M)$ and $M=M'$ in the case of pion
photoproduction.

\section*{APPENDIX B}

Let us now recall the implication of the EXD in the pairs,
$\pi$-$b_1$ and $\rho$-$a_2$. To determine the phase of the Regge
propagator, this notion together with
the $G$-parity consideration is of importance 
\cite{stro}. In the quark line diagram the $\gamma p\to \pi^+ n$
process is drawn by the uncrossed channel (planar diagram) for the
charge coupling nucleon exchange in the $s$-channel, whereas the
case of $\gamma n\to \pi^- n$ draws the crossed
channel(non-planar) for the $u$-channel exchange of the charge
coupling nucleon. The former diagram leads to a nonzero imaginary
part of the amplitude by the optical theorem. In the Regge pole
exchange, one possibility for this is to choose the phases of the
exchanged mesons in Eq. (\ref{born+}) to be rotating. But the
latter diagram leads the production amplitude to a real one,
which, then, means that all the phases are taken to be constant in
Eq. (\ref{born-}).

According to the finite energy sum rule (which states that the sum
of all resonances in low energy is, on the average, approximately
equal to the Regge pole at high energy), the dispersion integral
for the photoproduction amplitude is given by the sum of the
nonresonating background (BG) contribution and the resonance
contribution, each of which is related to the Pomeron (P) and the
Regge pole, respectively \cite{fesr},
\begin{eqnarray}
&&\int_{\nu_0}^{\bar\nu}d\nu\,\nu^n\, {\rm Im\,}{\cal
A}_{BG}(\nu,t)=\sum_{P}\gamma_P(t)\frac{\bar{\nu}^{\,\alpha_P+n+1}}{{\alpha_P}+n+1}\
, \label{fesr1}\\ &&\int_{\nu_0}^{\bar\nu}d\nu\,\nu^n\, {\rm
Im\,}{\cal
A}_{res}(\nu,t)=\sum_{Regge}\gamma_R(t)\frac{\bar{\nu}^{\,\alpha_R+n+1}}{{\alpha_R}+n+1}\
.\ \ \ \ \label{fesr}
\end{eqnarray}
The symmetric variable $\nu=\frac{s-u}{2M}$ and the $\gamma_P$ and
$\gamma_R$ are the respective residues for the Pomeron and Regge
pole exchanges. Since the Pomeron exchange is not allowed for the
charged pion case the duality in Eq. (\ref{fesr1}) is irrelevant.
Thus, the optical theorem for the $N^*$ resonances in the $\gamma
p\to \pi^+ n$ process implies that ${\rm Im\,}{\cal A}_{res}\neq
0$, so does the sum in the r.h.s. of Eq. (\ref{fesr}). This is the
case when $\gamma_{\pi}\neq-\gamma_{b_1}$, and
$\gamma_\rho\neq-\gamma_{a_2}$, meaning that the EXD in the pairs,
$\pi$-$b_1$ and $\rho$-$a_2$ are weak with the phases taken to be
imaginary (rotating), but with residues that differ from each
other in Eq. (\ref{born+}). For the $\gamma n\to \pi^- p$ case
where the amplitude is real by the quark diagram argument, the
${\rm Im\,}{\cal A}_{res}(\nu,t)=0$ in Eq. (\ref{fesr}) implies
$\gamma_{\pi}=-\gamma_{b_1}$, and $\gamma_\rho=-\gamma_{a_2}$, as
a consequence. From these duality arguments, we see that the
degree of the EXD is stronger in the $\gamma n\to \pi^-p$ process.

\section*{APPENDIX C}

The tensor meson dominance (TMD) states that the hadron
energy-momentum tensor current $\Theta^{\mu\nu}$ is dominated by
the tensor meson $T^{\mu\nu}$. In terms of the source-field
identity analogous to the VMD in Eq. (\ref{rho-decay}), this can
be written as \cite{raman}
\begin{eqnarray}\label{tmd-idd}
\Theta_{\mu\nu}=\frac{m_T^2}{g_T}T_{\mu\nu}\,,
\end{eqnarray}
for the source tensor having zero trace as well as zero divergence
for brevity. Then, the TMD in the hadron tensor current is
manifested by
\begin{eqnarray}\label{tmd}
<H|\Theta^{\mu\nu}(0)|H> =
\frac{<0|\Theta^{\mu\nu}|T><T,H|H>}{t-m_T^2}\,,
\end{eqnarray}
with the $H=\pi$, or $N$. The application of the TMD to the
energy-momentum tensor form factors of the pion, and of the
nucleon leads us to the universality of the tensor meson
couplings.

Let us now consider the $t$-channel exchange of the $f_2(1270)$
tensor meson in the $\pi N$ scattering with the $f_2'(1525)$
assumed decoupled. The interaction Lagrangian is given by
\begin{eqnarray}\label{vertex3}
&&{\cal
L}_{f_2\pi\pi}=\left(\frac{2g_{f_2\pi\pi}}{m_f}\right)\partial_\mu\pi\partial_\nu\pi\,
f_2^{\mu\nu} \
\end{eqnarray}
and the $f_2\pi\pi$ vertex is \cite{gold,renn}
\begin{eqnarray}\label{ver5}
<f_2,\pi|\pi>=\left(\frac{2g_{f_2\pi\pi}}{m_f}\right)2Q_\mu
Q_\nu\, \epsilon^{\mu\nu}_{(f)}
\end{eqnarray}
with $Q={1\over 2}(q+q')$. The decay width of $f_2\to \pi\pi$ is
empirically known and given by
\begin{eqnarray}\label{decay2}
\Gamma_{f_2\to \pi\pi}=\frac{\alpha}{60\pi}
\left(\frac{2g_{f_2\pi\pi }}{m_f}\right)^2\frac{1}{m_f^2} |k|^5\ ,
\end{eqnarray}
where $|k|={1\over 2\sqrt{s}}\sqrt{(s-(m+m')^2)(s-(m-m')^2)}$ is
the decay momentum and $\alpha$=6 for $f\pi\pi$ and $\alpha$=1 for
$a_2 KK$ \cite{giac}. From this we estimates $g_{f_2\pi\pi}=5.9$
from $\Gamma_{f_2\to \pi\pi}=156.9$ MeV and $g_{a_2 KK}=6.83$ from
$\Gamma_{a_2\to \pi\pi}=5.24$ MeV.

The energy-momentum tensor form factor of the pion is defined by
\begin{eqnarray}\label{meson}
<\pi(q')|\Theta^{\mu\nu}(0)|\pi(q)>=F_1^{\pi}(t)(q+q')^\mu
(q+q')^\nu \ ,
\end{eqnarray}
with the condition at $t=0$,
\begin{eqnarray}
F_1^\pi(0)=\frac{1}{2}\ ,
\end{eqnarray}
which is determined by the normalization of one pion state with
respect to energy \cite{ysoh}. Assume the $f_2$-pole dominance,
Eq. (\ref{tmd}), in the pion form factor in Eq. (\ref{meson}), and
use the source-field identity in Eq. (\ref{tmd-idd}) for the
$<0|\Theta^{\mu\nu}|f_2>$ together with the vertex function
$<f_2,\pi|\pi>$ in Eq. (\ref{ver5}) \cite{raman},
\begin{eqnarray}
&&<\pi(q')|\Theta^{\mu\nu}(0)|\pi(q)>\nonumber\\
&&=\frac{m_f^2}{g_f}\left(\frac{1}{m_f^2-t}\right)\left[\frac{g_{f_2\pi\pi
}}{m_f}(q+q')^\mu(q+q')^\nu\right]\,,
\end{eqnarray}
we obtain
\begin{eqnarray}
F_1^{\pi}(t)=\frac{m_f}{m_f^2-t}\frac{g_{f_2\pi\pi}}{g_f}\ ,
\end{eqnarray}
and at $t=0$, the universal coupling of $f_2$ meson is given by
\begin{eqnarray}
g_f=\frac{2g_{f_2\pi\pi}}{m_f}=\frac{11.8}{m_f}\ .
\end{eqnarray}

On the other hand, we write the nucleon energy-momentum tensor
form factors as
\begin{eqnarray}\label{}
&&<N(p')|\Theta^{\mu\nu}(0)|N(p)>=\bar{u}(p')\bigg\{\frac{1}{2}F_1(t)(\gamma^\mu
P^\nu+\gamma^\nu P^\mu)\nonumber\\&&+\frac{F_2(t)}{M} P^\mu P^\nu
\bigg\}u(p),
\end{eqnarray}
with the following conditions at $t=0$,
\begin{eqnarray}\label{norm}
F_1(0)+F_2(0)=1\,, \ \ \ F_2(0)=0\,,
\end{eqnarray}
which are determined from the normalization of one nucleon state
with energy and spin $\frac{1}{2}$. In a similar fashion, the
$f_2$ meson dominance, Eq. (\ref{tmd}), in these form factors
leads to the following expression,
\begin{widetext}
\begin{eqnarray}
<N(p')|\Theta^{\mu\nu}(0)|N(p)>=\frac{m_f^2}{g_f}\left(\frac{1}{m_f^2-t}\right)
\bar{u}(p')\bigg\{\frac{2g^{(1)}_{f_2NN}}{M}(\gamma^\mu
P^\nu+\gamma^\nu P^\mu)+\frac{4g^{(2)}_{f_2NN}}{M^2} P^\mu
P^\nu\bigg\} u(p)\,,
\end{eqnarray}
\end{widetext}
where we use Eq. (\ref{ten00}) for the vertex $<f_2,N|N>$. Thus,
the TMD in the nucleon energy-momentum form factors are given by
\begin{eqnarray}\label{stress}
&&F_1(t)=\frac{m_f^2}{m_f^2-t}\frac{4g^{(1)}_{f_2NN}}{g_f M}\
,\nonumber\\
&&F_2(t)=\frac{m_f^2}{m_f^2-t}\frac{4g^{(2)}_{f_2NN}}{g_fM}\ .
\end{eqnarray}
Combining the above equations with the conditions, Eq.
(\ref{norm}) at $t=0$, we get the following identities,
\begin{eqnarray}\label{tmd01}
\frac{2}{M}(g^{(1)}_{f_{2NN}}+g^{(2)}_{f_{2NN}})=\frac{g_{f_2\pi\pi
}}{m_f}\,, \ \ \  g^{(2)}_{f_{2NN}}=0\,,
\end{eqnarray}
and the universality of coupling constant $g_f$ is given in the
form,
\begin{eqnarray} \frac{4g^{(1)}_{f_2 NN}}{M}=g_f=\frac{2g_{f_2\pi\pi
}}{m_f}\ .
\end{eqnarray}
For the $KN$ scattering we obtain the same identities as in Eq.
(\ref{tmd01}) which relate the $g^{(1)}_{a_{2NN}}$ and
$g^{(2)}_{a_{2NN}}$ with $g_{a_2 K\bar{K}}$ by virtue of the TMD.

\end{document}